\documentclass[twocolumn]{aastex631}
\usepackage{graphicx} 

\usepackage{amsmath}

\newcommand{\mcl}{$M_{\text{CL}}$}

\newcommand{\frun}{$f_{\rm{run}}$}

\def\code#1{\texttt{#1}}

\begin{document}

\title{Binary black hole mergers and intermediate-mass black holes in dense star clusters with collisional runaways}

\correspondingauthor{Rujuta A. Purohit}
\email{rujuta.purohit.24@dartmouth.edu}

\author[0009-0003-6728-9291]{Rujuta A. Purohit} 
\affil{Department of Physics and Astronomy, Dartmouth College, 6127 Wilder Laboratory, Hanover, NH 03755, USA}
\affiliation{Center for Interdisciplinary Exploration \& Research in Astrophysics (CIERA), Northwestern University, Evanston, IL 60208, USA}
\author[0000-0002-7330-027X]{Giacomo Fragione}
\affiliation{Center for Interdisciplinary Exploration \& Research in Astrophysics (CIERA), Northwestern University, Evanston, IL 60208, USA}
\affiliation{Department of Physics \& Astronomy, Northwestern University, Evanston, IL 60208, USA}
\author[0000-0002-7132-418X]{Frederic A. Rasio}
\affiliation{Center for Interdisciplinary Exploration \& Research in Astrophysics (CIERA), Northwestern University, Evanston, IL 60208, USA}
\affiliation{Department of Physics \& Astronomy, Northwestern University, Evanston, IL 60208, USA}
\author[0000-0001-6941-8411]{Grayson C. Petter}
\affil{Department of Physics and Astronomy, Dartmouth College, 6127 Wilder Laboratory, Hanover, NH 03755, USA}
\author[0000-0003-1468-9526]{Ryan C. Hickox}
\affil{Department of Physics and Astronomy, Dartmouth College, 6127 Wilder Laboratory, Hanover, NH 03755, USA}

\shorttitle{BBH mergers and IMBHs in star clusters with runaways}

\shortauthors{Purohit et. al}

\begin{abstract}
    Intermediate-mass black holes (IMBHs) are believed to be the missing link between the supermassive black holes (BHs) found at the centers of massive galaxies and BHs formed through stellar core collapse. One of the proposed mechanisms for their formation is a collisional runaway process in high-density young star clusters, where an unusually massive object forms through repeated stellar collisions and mergers, eventually collapsing to form an IMBH. This seed IMBH could then grow further through binary mergers with other stellar-mass BHs. Here we investigate the gravitational-wave (GW) signals produced during these later IMBH--BH mergers. We use a state-of-the-art semi-analytic approach to study the stellar dynamics and to characterize the rates and properties of IMBH--BH mergers. We also study the prospects for detection of these mergers by current and future GW observatories, both space-based (LISA) and ground-based (LIGO Voyager, Einstein Telescope, and Cosmic Explorer). We find that most of the merger signals could be detected, with some of them being multi-band sources. Therefore, GWs represent a unique tool to test the collisional runaway scenario and to constrain the population of dynamically assembled IMBHs.
\end{abstract}


\section{Introduction}
\label{sec:intro}

Black holes (BHs) are observed commonly in two mass ranges: stellar-mass BHs, formed through massive star evolution to core collapse, with masses $M_{\rm{BH}} \lesssim 100\,M_{\odot}$ \citep[e.g.,][]{Celotti1999, Remillard2006,TheLIGOScientificCollaborationtheVirgoCollaboration2021}, and supermassive BHs found at the centers of most massive galaxies, with $M_{\rm{BH}} \gtrsim 10^5\,M_{\odot}$ \citep[e.g.,][]{Tremaine2002, Marconi2003, Kormendy2013}. BHs with masses in between these two regimes are labeled intermediate-mass BHs (IMBHs); for a review see \citet{Greene2020}. While IMBHs could play a fundamental role in the evolution of galaxies and could be a source of tidal disruption events and gravitational waves (GWs), their existence has not been confirmed beyond a reasonable doubt \citep[e.g.,][]{Jardel2012, Neumayer2012, GrahamScott2013, Mezcua2017, Nguyen2018, PerleyMazzali2019, SmithMagno2023}.

IMBHs have masses beyond the most massive BH that can be produced as a result of direct stellar core collapse. Current stellar evolution models predict a dearth of BHs with masses larger than about $50\,M_{\odot}$ as a result of pair-instability physics, where pair production removes pressure support in the core \citep[][]{Heger2003, Woosley2017}. Whenever the pre-explosion stellar core is in the mass range $M_{\star} \sim 45-65\,M_{\odot}$ at the onset of their carbon burning, large amounts of mass can be ejected, leaving a BH remnant with a maximum mass around $50\,M_{\odot}$. Larger stellar cores can trigger thermonuclear reactions that can completely destroy the star and leave no remnant behind.

Several mechanisms have been proposed for the formation of IMBHs. These include direct collapse of primordial gas clouds of pristine gas \citep[e.g.,][]{Bromm2003, Begelman2006}, the remnants of massive Population III stars ($140 \lesssim M_{\star} \lesssim 260 M_{\odot}$) in the early Universe \citep[e.g.,][]{Madau2001, Whalen2012, Jiang2019}, repeated mergers of main-sequence stars later collapsing into a massive BHs \citep[e.g.,][]{Portegies2002, Gurkan2004, GierszLeigh2015, DiCarlo2021, Gonzalez2021}, or hierarchical mergers of stellar-mass BHs \citep[e.g.,][]{Miller2002, Antonini2019, Fragione2022}.

IMBHs are primary sources for present and future GW observatories. Recent detections by the LIGO-Virgo-KAGRA (LVK) Collaboration have found binary BH (BBH) mergers where one or both components of the merging binary have masses above $50\, M_{\odot}$. Among them, GW190521 is the most interesting event since its remnant has a total mass of $\sim 150\, M_{\odot}$, nominally in the IMBH regime \citep{Abbott2020}. One of the main venues for the assembly of these binaries is the core of dense star clusters, where a massive BH could form through collisions and mergers of massive stars, and through hierarchical BH mergers \citep{Antonini2016, Gonzalez2021, Chattopadhyay2023, FragioneRasio2023, Atallah2023}. With hundreds of new detections expected in the current O4 run by the LVK Collaboration, studying BBH mergers across all relevant mass ranges has become a key priority.

In this paper, we model repeated mergers of BBHs in dense star clusters that we assume to have undergone at early times (typically within just a few Myr after cluster formation) a collisional runaway process leading to the formation of a very massive star (with $M_{\star} \sim 10^2-10^3\,M_\odot$) and, ultimately an IMBH. We perform simulations of merging BBHs formed in these dense clusters using the semi-analytic method developed in \citet{FragioneRasio2023}. The remnants of the merging BBHs grow the IMBH to final masses $\geq 10^3 M_{\odot}$. We explore for the first time the implications of such a runaway process for GW detections by present and future observatories.

Our paper is organized as follows: in Section~\ref{sec:methods} we discuss our semi-analytic method to study BH mergers in dense star clusters hosting an IMBH. In Section~\ref{sec:results} we present our numerical results, and we summarize our key findings and implications for future work in Section~\ref{sec:conc}.

\section{Methods}
\label{sec:methods}

In this section, we include a summary of the semi-analytic approach we use to derive our results. For a detailed discussion, see \S\,2 of \citet{FragioneRasio2023}.

We consider a dense stellar cluster of initial mass \mcl, in the range $[10^5, \, 10^7] \; M_{\odot}$, and half-mass–radius $r_{\rm h} = 1$ \, pc, described by a King model with initial moderate concentration \citep{King1962}. This is representative of the typical size of a young massive cluster \citep{Portegies2010}. We sample stellar masses from the canonical Kroupa initial mass function \citep{Kroupa2001}:
\begin{equation}
    \xi(m_{\star}) \propto
    \begin{cases}
        \left(\frac{m_{\star}}{0.5 M_{\odot}}\right)^{-1.3} & \text{if } 0.08 \leq \frac{m_{\star}}{M_{\odot}} \leq 0.50 \\
        \left(\frac{m_{\star}}{0.5 M_{\odot}}\right)^{-2.3} & \text{if } 0.50 \leq \frac{m_{\star}}{M_{\odot}} \leq 150\,.
    \end{cases}
\end{equation}
Given this IMF, we produce a total of
\begin{equation}
    N_{\rm{BH}} = 3.025 \times 10^3 \left(\frac{M_{\rm{CL}}}{10^6 M_{\odot}} \right)
\end{equation}
BH progenitors, in the mass range $[20 M_{\odot}, \, 150 M_{\odot}]$. We consider cluster metallicities $Z \in \{0.02, 0.002, 0.0002\}$, and evolve the BH progenitors at a given metallicity using the stellar evolution code \textsc{sse} \citep{Hurley2000}, with all the necessary updated prescriptions for stellar winds, stellar interactions, and formation of remnants \citep[for details see][]{2020Banerjee}. 

We assume that all BHs are born with negligible natal spins, consistent with the findings of  \citet{FullerMa2019}. In our simulations, each BH is imparted a natal kick at birth sampled from a Maxwellian distribution with velocity dispersion $265\,\rm km\,s^{-1}$ \citep{Hobbs2005}, scaled by a factor of $1.4 M_{\odot}/M_{\rm{BH}}$ to account for momentum conservation \citep{FryerKalogera2001}. If the natal kick exceeds the cluster escape velocity \citep{FragioneRasio2023}
\begin{equation}
    v_{\rm {esc}} = 32 \, {\rm{km\,s^{-1}}} \, \left(\frac{M_{\rm {CL}}}{10^5 \, M_{\odot}} \right)^{1/2} \left(\frac{r_{\rm{h}}}{\rm{1 \, pc}} \right)^{-1/2}\,,
\end{equation}
we consider the BH ejected and remove it from our simulation. The same applies for dynamical kicks in three-body encounters and recoil kicks imparted to the remnant of a BBH merger, caused by anisotropic GW emission \citep[see e.g.,][]{LoustoCampanelli2010}. 

We model cluster evolution by following \citet{Antonini2020a, Antonini2020b}. Briefly, the cluster is assumed to reach a state of balanced evolution, so that the heat generated by the BBHs in the core and the cluster global properties are related. Our simulations include all the fundamental elements of cluster evolution (cluster mass loss and expansion) and the fundamental processes of formation and evolution of BBHs (3-body interactions, mergers, recoil kicks, etc.). In \citet{FragioneRasio2023}, it was shown that this semi-analytic method performs well at reproducing the essential elements of BBH mergers, especially when compared to $N$-body simulations. For more details, see \citet{Antonini2020a, Antonini2020b} and \citet{FragioneRasio2023}.

In our study, we build on the scheme presented by \citet{FragioneRasio2023} by adding a new parameter $f_{\rm{run}}$, which represents the fraction of total cluster mass that participates in the initial runaway process. The mass of the star formed as a result of the runaway is then simply
\begin{equation}
    m_{\text{run}} = f_{\rm run}\,M_{\rm CL}.
\end{equation} 
Many previous works have tried to estimate \frun from Monte Carlo N-body simulations, finding a typical value of $\sim 10^{-3}$ \citep{Freitag2001, Gurkan2004}. We assume that every cluster goes through a runaway process initially, regardless of its initial mass and density. Note that however runaways typically develop in clusters with high densities, with other parameters such as the slope of the initial mass function, primordial binary fraction in massive stars, etc. that could play a critical role \citep[e.g.,][]{Fregeau2002, Ivanova2005, Gonzalez2021}. 

In our simulations, we consider four different cases, with $f_{\rm run} \in \{0, 0.0005, 0.001, 0.005, 0.01\}$. For each value of $f_{\rm run}$, we run 5000 simulations for each metallicity. The clusters have masses in the range \mcl $\in [10^5, 10^7] \; M_{\odot}$ assuming a distribution of cluster masses $\propto M_{\rm CL}^{-2}$ \citep{Portegies2010}.

With our simulations, our goal is to measure the rates of BBH mergers for BHs formed through direct collapse of large stars, study the population of the merger remnants by tracking the growth of IMBHs, and the probability of the detection of these merger events using GW instruments.

\section{Results}
\label{sec:results}

In this Section, we summarize the results of our simulations and discuss the detectability of the BBH mergers by present and upcoming GW observatories.

\subsection{Growth of an IMBH}

\begin{figure}[h]
    \centering
    \includegraphics[width=0.45\textwidth]{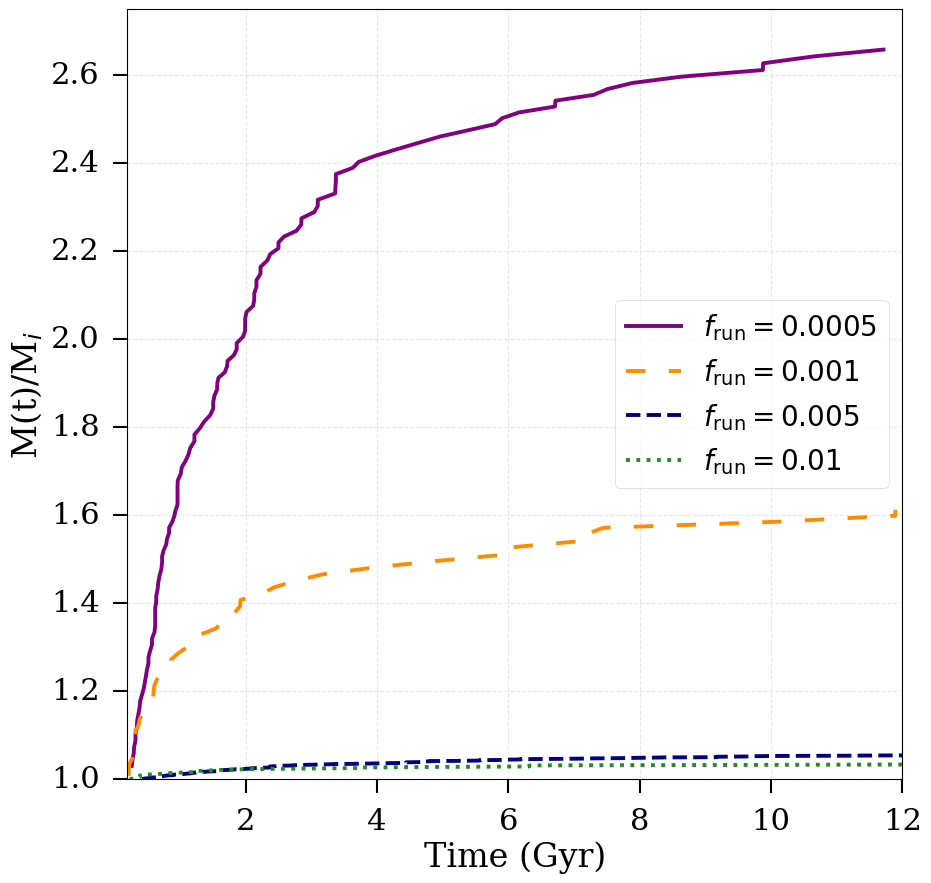}
    \caption{The growth of IMBHs (ratio of its mass to the initial mass) in clusters with initial cluster mass \mcl $= 10^6 M_{\odot}$ as a function of time for different values of \frun. For clusters with higher values of \frun, the ratio does not increase significantly, while IMBHs in clusters with smaller values of \frun \, grow significantly over time.}
    \label{fig:imbh_time}
\end{figure}

\begin{figure}[t]
    \centering
    \includegraphics[width=0.45\textwidth]{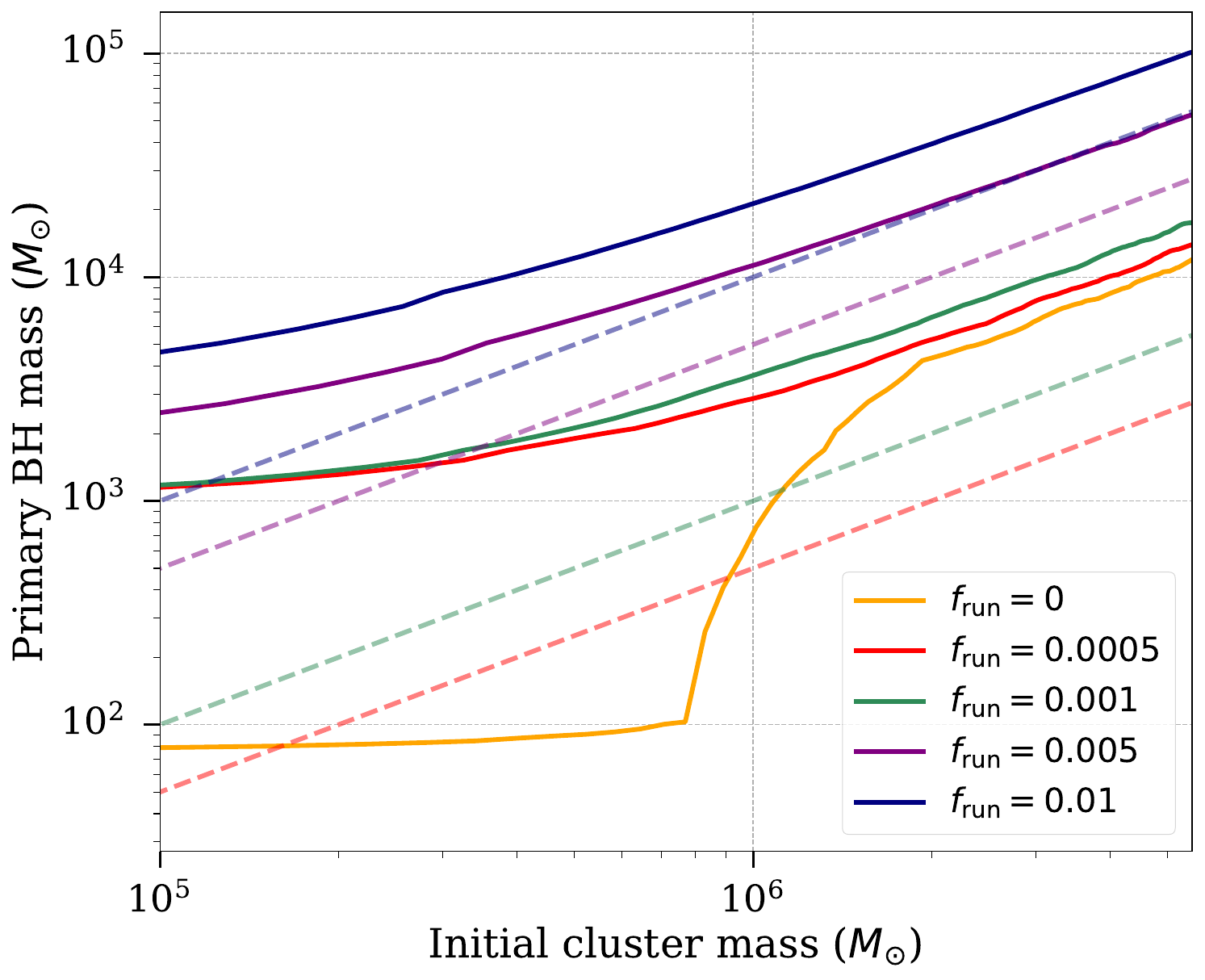}
    \caption{Solid lines: most massive BH formed in a cluster as a function of the initial cluster mass \mcl \, for different values of the fraction of cluster mass that undergoes initial collisional runaway, $f_{\rm run}$. The half-mass radius is fixed at $r_{\rm h}=1$\,pc representative of the typical size of young massive clusters. Dashed lines: initial IMBH mass, $m_{\rm{run}} = f_{\rm{run}} M_{\rm{CL}}$, as a function of \mcl. The simulations contain three metallicities as $Z \in \{0.02, 0.002, 0.0002\}$.}
    \label{fig:massive_bh}
\end{figure}

\begin{figure*}[t]
    \centering
    \includegraphics[width=\textwidth]{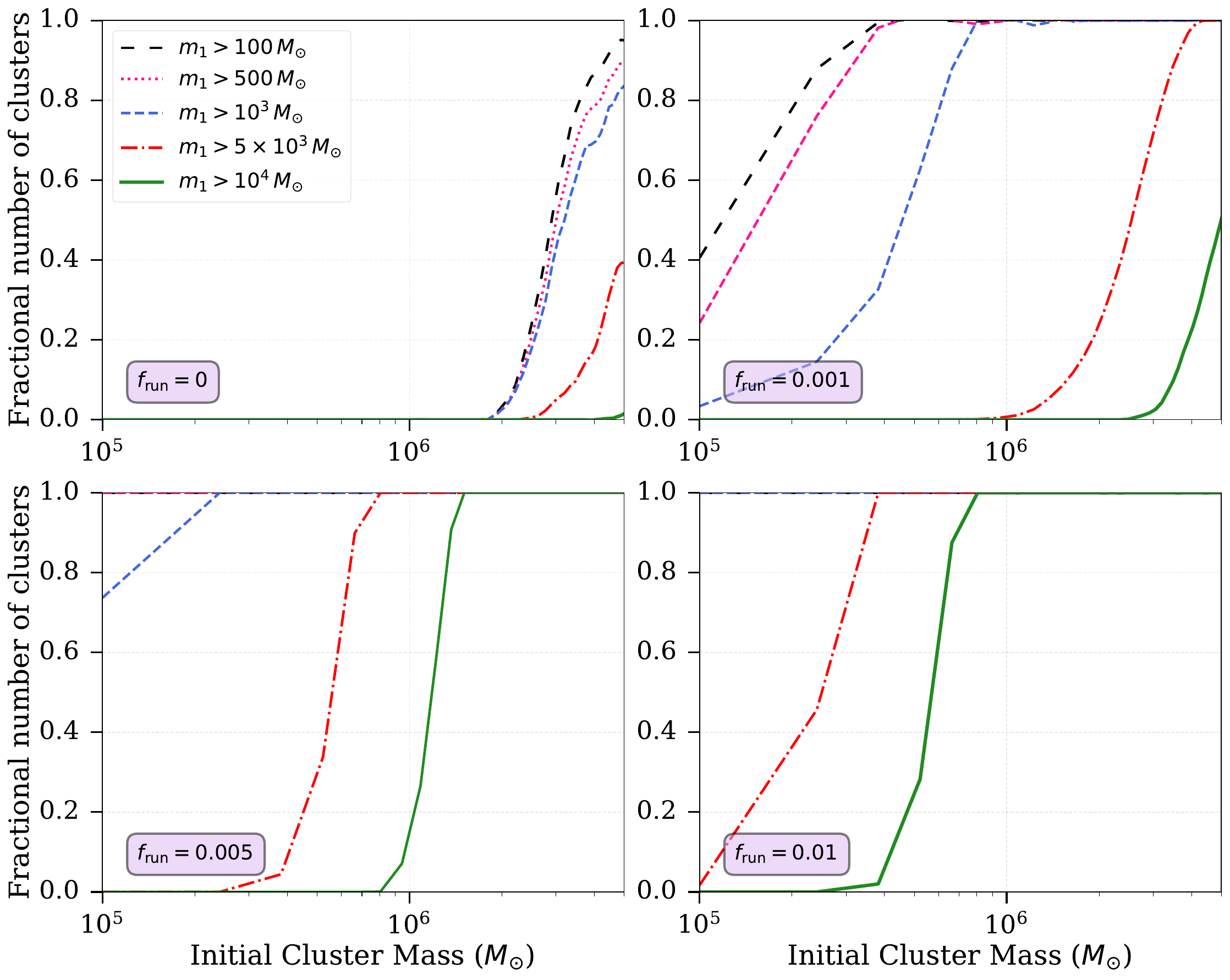}
    \caption{The fraction of clusters that form a massive BH as a function of the initial cluster mass. As expected for larger values of \frun, this fraction increases, especially for lower cluster mass. We show 5 different mass thresholds for what we call a massive BH: $100, \, 500, \, 10^3, \,5 \times 10^3, \, 10^4 M_{\odot}$.}
    \label{fig:numcl}
\end{figure*}

In our simulations, we initialized our star clusters with a massive BH remnant of the collapse of a large star resulting from the runaway process \citep[e.g.,][]{Portegies2002, Zwart2004, Gurkan2006, Giersz2015, Mapeli2016, Gonzalez2021}. This BH may grow to an IMBH through repeated mergers with other BHs over time, whenever not ejected as a consequence of dynamical kicks in few-body interactions or recoil kicks after a merger via GW emission.

We track the growth of the most massive BH in each cluster as a result of repeated mergers with other stellar BHs. In Figure~\ref{fig:imbh_time}, we show the growth of IMBHs (ratio of its mass to the initial mass) in clusters with initial cluster mass \mcl $= 10^6 M_{\odot}$ as a function of time for different values of \frun. For clusters with higher values of \frun, the ratio does not increase significantly; IMBHs in clusters with smaller values of \frun \, grow significantly over time, reaching 2-3 times their initial mass.

In Figure \ref{fig:massive_bh}, we show the most massive BH formed in a star cluster as a function of the initial cluster mass for various values of the fraction of cluster mass that undergoes a runaway process. For simulations sampling all three metallicities, we make bins of cluster masses and select the most massive BH in each bin. As expected, a larger value of $f_{\rm run}$ implies a larger most massive BH across all initial cluster masses. For example, we find that a cluster of initial mass of $10^5\,M_\odot$, \frun$ = 0$ leads to a most massive BH of about $80 M_{\odot}$, \frun$ = 0.001$ of about $1.1 \times 10^3 M_{\odot}$, and \frun$ = 0.01$ of about $ 4.5 \times 10^3 M_{\odot}$.

The value of $f_{\rm run}$ is particularly crucial for small clusters. In particular, this is because for smaller values of \frun \, the smaller IMBHs are ejected from the cluster due to dynamical or recoil kicks after the mergers. For example, in the case there is no runaway ($f_{\rm run}=0$), clusters with masses $\lesssim 10^6\,M_\odot$ do not form an IMBH; however, the most massive IMBH for a cluster of mass $\sim 10^6 M_{\odot}$ has a mass of about $5 \times 10^3 M_{\odot}$ for \frun $=0.001$, while a mass of about $3 \times 10^4 M_{\odot}$ for \frun $=0.01$.

To provide additional insight into the formation of massive BHs within clusters, Figure~\ref{fig:numcl} illustrates the fraction of clusters, or \textit{likelihood} of the most massive BH (referred to as $m_1$ in Figure~\ref{fig:numcl}) surpassing a specified mass threshold in our simulations. When \frun$ = 0$ (no runaway), only very massive clusters exhibit a non-negligible probability of forming massive BHs. Conversely, with higher \frun \ values, the initial mass of the IMBH is larger, increasing the likelihood of their appearance even in less massive clusters, as ejections also become less important.\\

\newpage

\subsection{BBH merger rates}
\label{sec:rates}

\begin{figure*}[t]
    \centering
    \includegraphics[width=\textwidth]{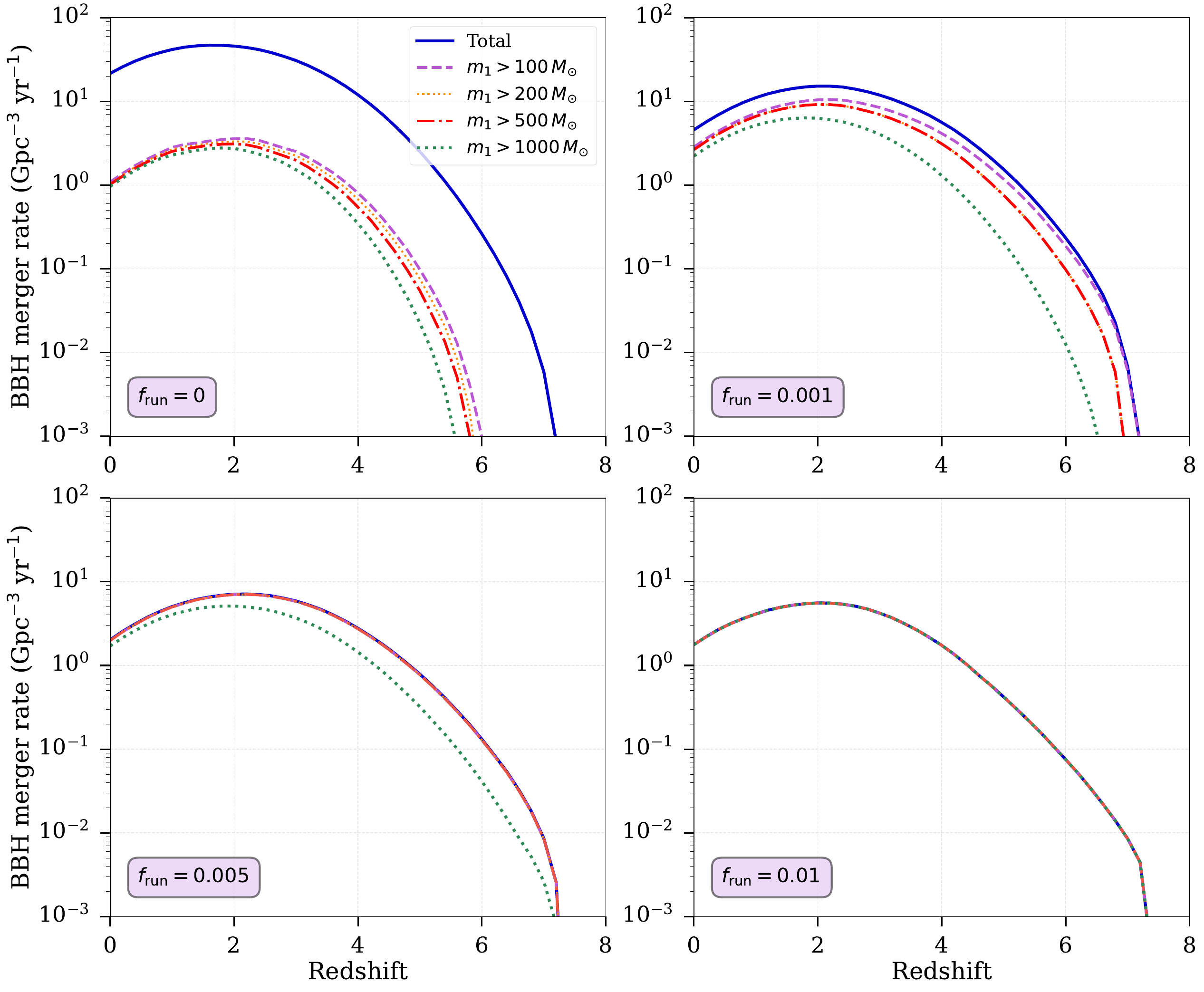}
    \caption{The binary black hole (BBH) merger rate as a function of redshift for 4 different fractions of initial cluster mass that undergo the collisional runaway. The total merger rate is shown, along with rates for a primary mass over a given threshold for the primary mass: $100, 200, 500, 1000 M_{\odot}$.}
    \label{fig:rates}
\end{figure*}

We follow \citet{FragioneRasio2023} and calculate the rates (in units of Gpc$^{-3}$ yr$^{-1}$) of BBH mergers as\footnote{In this work, we adopt a standard $\Lambda$CDM cosmology \citep{Planck2016}: $H_0 = 67.8$ km s$^{-1}$ Mpc$^{-1}$, $\Omega_{\text{m}} = 0.308$, and $\Omega_{\Lambda} = 0.692$.}
\begin{equation}
    \begin{aligned}
    R(z) &= K \frac{d}{dt_{\text{lb}}} \int \int \int \int dM_{\text{CL}} \, dr_h \, dZ \, dz_f \, \frac{dt_{\text{lb}}}{dz_f} \\
    &\quad \times \frac{\partial N_{\text{events}}}{\partial M_{\text{CL}} \, \partial r_h \, \partial Z \, \partial z_f} \Psi(M_{\mathrm{CL}} r_h, Z, z_f)\,,
    \end{aligned}
    \label{eq:rates}
\end{equation}
where $t_{\rm{lb}}$ is the look-back time at redshift $z$, $N_{\rm{events}}$ is the number of events as a function of the initial cluster mass $M_{\rm CL}$, initial half-mass radius $r_{\rm h}$, metallicity $Z$, and formation redshift $z_{\rm f}$, and $\Psi(M_{\rm{CL}}, r_h, Z, z_f)$ is a probability function that weighs the the previous cluster properties. In our model, we take the distribution of cluster masses to be $\propto M_{\rm CL}^{-2}$, with the maximum possible cluster mass being \mcl $ = 10^7 M_{\odot}$ \citep[e.g.,][]{Portegies2000}. As discussed above, we fix the half-mass radius at $r_{\rm h}=1$\,pc, which follows the typical distribution of observed values for local, young stellar clusters \citep{Portegies2010}. We take the formation times to be proportional to $\exp[-(z - z_f)^2/(2 \sigma_f^2)]$ where $z_f = 3.2$ and $\sigma_f = 1.5$, reminiscent of cluster formation times as inferred from cosmological simulations \citep{Mapelli2021}. The cluster masses normalized such that the cluster density is $\sim 1$ Mpc$^{-3}$ in the local Universe \citet{Portegies2010}. Metallicities are sampled from a log-normal distribution with mean given by \citep{MadauFragos2017}
\begin{equation}
\log \langle Z/{\rm Z}_\odot \rangle = 0.153 - 0.074\,z^{1.34}
\end{equation}
and a standard deviation of $0.5$~dex. In Equation \ref{eq:rates}, $K$ accounts for the cluster density evolution, considering that a fraction of the star clusters that are formed in the Universe evaporate across cosmic time. We fix $K = 32.5$, consistent with \citet{Antonini2020a} and with the value needed to reproduce the merger rates of BBH mergers in the latest LVK catalog \citep{Fishbach2023}.

We calculate the BBH merger rate as a function of redshift for each values of the runaway fraction. Figure~\ref{fig:rates} shows the overall merger rate for our models, which we also break down to show the contribution of IMBH of different masses ($m_1 > 100, 200, 500, 1000\,M_{\odot}$). We find that relative contribution of massive BHs is larger for larger values of \frun, since the IMBH starts dominating the merger rate, as it keeps merging repeatedly with the other stellar-mass BHs. Indeed, we find that \frun = 0 has a merger rate of 19 Gpc$^{-3}$ yr$^{-1}$, \frun = 0.001 of 6.2 Gpc$^{-3}$ yr$^{-1}$, \frun = 0.005 of 2.9 Gpc$^{-3}$ yr$^{-1}$, and \frun = 0.01 of 2.24 Gpc$^{-3}$ yr$^{-1}$. At the same time, the merger rate for primary masses larger than $1000 M_{\odot}$ for \frun = 0 is of 0.09 Gpc$^{-3}$ yr$^{-1}$, for \frun = 0.001 of 1.66 Gpc$^{-3}$ yr$^{-1}$, for \frun = 0.005 of 2.03 Gpc$^{-3}$ yr$^{-1}$, and for \frun = 0.01 of 2.24 Gpc$^{-3}$ yr$^{-1}$.

\subsection{Detectability of merging BBH with GW instruments}

\begin{figure*}
    \centering
    \includegraphics[width=0.85\textwidth]{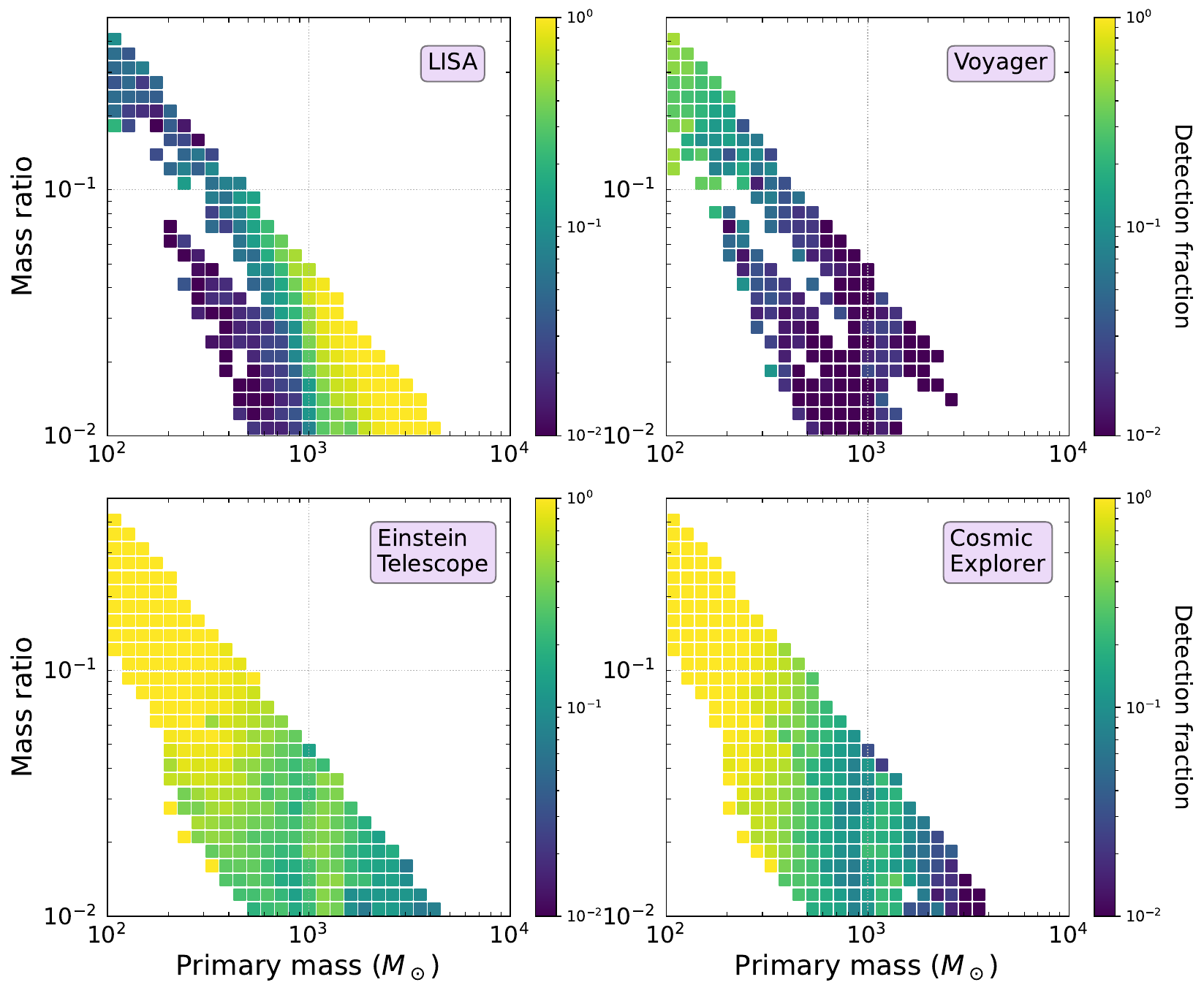}
    \caption{Detection fractions for merging IMBH binaries as a function of primary mass and binary mass ratio.
    This figure shows our results for clusters with \frun = 0.001 (see Appendix for other values). Top left: LISA; top right: Voyager; bottom left: Einstein Telescope; bottom right: Cosmic Explorer.}
    \label{fig:gw_001}
\end{figure*}

Given a population of merging BBHs, we must ask whether current or future ground-based and space-based instruments are likely or not to detect some of them. We calculate the detection fraction of these merger events for planned ground- and space-based GW instruments, such as the Laser Interferometer Space Antenna \citep[LISA;][]{Robson2019}, Voyager \citep[the upgraded detector of the current LIGO facility,][]{2019LIGOfuture}, the Einstein Telescope \citep[ET;][]{Punturo2010}, and the Cosmic Explorer \citep[CE;][]{Reitze2019}. While LVK could detect the mergers of BBHs with $\sim 100 M_{\odot}$ out to redshifts of $z \sim 1$ \citep[][]{Abbott2019}, the upcoming space-based LISA and ground-based instruments like ET and CE offer the opportunity of detecting the formation of IMBHs up to $z \sim 100$ \citep[][]{Amaro2017, Jani2020, FragioneLoeb2023}. Thus, understanding the probability of detection of these events will guide searches for massive BBH systems and help plan future observing cycles.

Following \citet{FragioneLoeb2023}, we design the detection fraction as an instrument-dependent function, which encodes the ability of observing the merger of a binary with primary mass $m_1$ and mass ratio \textit{q} at a redshift \textit{z}. The masses and merger times are obtained directly form our simulations. However, the merger times need to be corrected for the cluster formation time. As for the merger rates, we take the formation times to be proportional to $\exp[-(z - z_f)^2/(2 \sigma_f^2)]$ where $z_f = 3.2$ and $\sigma_f = 1.5$ described in \S \ref{sec:rates}. Once we correct for the cluster formation, we discard those binaries that have a merger time larger than the age of the Universe. The sources not discarded are kept in our data and their merger times are converted to redshifts. 

The detection of a merger event is modeled as an instrument-dependent function $ F_{\mathrm{det}}(z, m_1, q)$ as
\begin{equation}
    F_{\mathrm{det}}(z, m_1, q) = H(\langle\rho(z, m_1, q)\rangle >\rho_{\mathrm{thresh}})\,,
\end{equation}
where $H$ is the Heaviside function and  $\langle\rho(z, m_1, q)\rangle$ is the averaged signal-to-noise (SNR) ratio. The threshold for a detection is set to an SNR of 8. From \citet{FragioneLoeb2023}, we compute the average SNR as
\begin{equation}
    \langle\rho(z, m_1, q)\rangle =   2C \ \sqrt{\int_{f_{\mathrm{min}}}^{f_{\mathrm{max}}} \frac{|\Tilde{h}{f}|^2}{S_n(f)} df}\,,
\end{equation}
where $C$ is obtained after averaging over various sky relations, with $C = 2/\sqrt{5}$ and $C = 2/5$ for space-based and ground-based detectors, respectively \citep{Robson2019}; $f_{\rm{min}}$ and $f_{\rm{max}}$ are the minimum and maximum frequency of the binary in the detector band, respectively; $S_{\rm{n}}(f)$ is the noise power spectral density; $|\Tilde{h}(f)|$ is the frequency-domain waveform amplitude for a face-on binary. We use pyCBC developed by \citet{Nitz2019} with the IMRPhenomD approximant \citep{Husa2016} to model the waveform of the merging binary BHs.

We study the detection probability of these merger events using four different gravitational wave observatories: space-based LISA and ground-based Voyager, ET, and CE. The power spectral density of LISA is derived as in \citet{Robson2019}, of Voyager as in \citet{2019LIGOfuture}, of ET as in \citet{Punturo2010}, and of CE as in \citet{Reitze2019}. LISA has a planned duration of 5 years, and a binary will evolve towards a merger event as it starts from the frequency $f_{\rm{in}}$ which is the GW frequency at the start of the observation
\begin{equation}
    \begin{aligned}
        f_{\rm{ini}} &= 1.2 \times 10^{-2} \, \rm{Hz} \, (1 + z)^{-\frac{5}{8}} \, \left(\frac{1 + q}{q^3} \right)^{\frac{1}{8}}\\
        &\times \left(\frac{m_1}{100 M_{\odot}} \right)^{-\frac{5}{8}} \, \left(\frac{T_{\rm{LISA}}}{4 \rm{yr}} \right)^{-\frac{3}{8}}\,.
    \end{aligned}
\end{equation}
This initial frequency is higher than the minimum detectable frequency for LISA which is conventionally $10^{-5}$ Hz. Thus, we take, $f_{\rm{min}} = f_{\rm{ini}}$ for LISA. For the ground-based instruments in the study namely Voyager, ET, and CE, this initial frequency $f_{\rm{ini}}$ is typically smaller than the frequency at which the instruments start operating, 5 Hz, 1 Hz, and 5 Hz, respectively.

In Figure \ref{fig:gw_001}, we show the detection probability of BBH merger events in clusters, with \frun = 0.001. We see that LISA is able to detect most of the mergers with primary mass $\gtrsim 10^3 M_{\odot}$ and mass ratio $10^{-2} \lesssim q \lesssim 10^{-1}$. For the smaller masses, LISA is only able to detect a few of the merging IMBH binaries. Conversely, the population of merging binaries with $m_1 \lesssim 10^3 M_{\odot}$ can be studied using ground-based instruments. LIGO's Voyager is only able to detect a few of the merging binaries in this regime, in particular when the mass ratio is close to unity. ET and CE are able to detect most binaries with primary mass $m_1 \lesssim 10^3 M_{\odot}$ and mass ratio $\sim 10^{-1}$.

The plots showing detection fractions for other values of \frun \, are included in the Appendix.

\section{Conclusion and Discussion}
\label{sec:conc}

In this paper, we analyzed simulations of merging BBHs in dense clusters, where an IMBH is assumed to have formed as result of the collapse of a very massive star created from repeated stellar mergers in a collisional runaway. We can summarize our key findings as follows:
\begin{itemize}
    \item The initial IMBH, born as the remnant of a very massive star formed through the collisional runaway, grows in time through repeated mergers with stellar-mass BHs, up to about 2--3 times its initial mass for small values of $f_{\rm run}$ (the fraction of initial cluster mass that goes into the collisional runaway).
    \item Through collisional runaways star clusters may produce IMBHs with initial masses $\sim 10^2$-$10^3\,M_{\odot}$, which further grow to $\gtrsim 10^3 M_{\odot}$ through repeated mergers with other stellar-mass BHs. This is not seen in any of our models when we set \frun = 0, i.e., when BHs grow solely through BH mergers.
    \item The merger rate of BBHs dynamically assembled in dense star clusters tends to decrease if the mass of the IMBH formed as a result of a runaway process is larger. For sufficiently large masses, the IMBH dominates the merger rate and the merger rate of stellar-mass BHs becomes negligible.
    \item The merging binaries that LISA can potentially detect map the underlying astrophysical population for primary masses $m_1 > 100\,M_{\odot}$ and mass ratios $1 > q > 10^{-2}$. The ground-based instruments ET and CE will be capable of observing the higher mass ratio binaries, with primary mass $m_1 < 10^3 M_{\odot}$ and mass ratio $q > 10^{-2}$, while the detection efficiency of LIGO Voyager is typically very small.
\end{itemize}

Our work considered a potentially important effect in the formation of massive BHs and their mergers that previous models have neglected. However, there are some caveats that we leave to future work. For example, we have not included in our treatment the interaction of the IMBH with ordinary stars, which could lead to interesting phenomena, such as tidal disruption events \citep[e.g.,][]{Liu2009, Fragione2021, 2022Angus, Fulya2023}. We have also fixed some of the distributions that describe the birth properties of star clusters, consistent with observed properties in the local universe \citep{Portegies2010}, but these are poorly constrained elsewhere. Finally, clusters with low densities are less likely produce runaways, while we assume that all clusters undergo an initial runaway.

Furthermore, we have used the IMRPhenomD as our approximation for the waveform, which only includes the dominant harmonic $(l,m) = (2, 2)$ of the GW signal. However, higher-order harmonics could contribute to the GW signal significantly, especially for IMBH binaries \citep{Jani2020}. Finally, another important yet currently uncertain question is the fate of the very massive star (with mass $\sim 10^2-10^3\,M_\odot$) that is produced as result of the runaway process, as stellar evolutionary models have not been calibrated over that mass range. As \citet{Mapeli2016} notes, the mass of the final remnant of the massive star is in the IMBH mass range only if the mass loss due to stellar winds or hydrodynamic processes is moderate, and only if the very massive star undergoes direct collapse to a BH. These conditions impose many new restrictions.

GW detection offers an unparalleled opportunity to survey the Universe and detect IMBHs in various mass ranges, making it possible for the first time to constrain their formation, growth, and merger history across cosmic time.

From our analysis of merging BH--IMBH binaries in dense star clusters, we find that the merger rates are $\sim 10$ Gpc$^{-3}$ yr$^{-1}$ and peak around redshift $z \simeq 2$. Thus, we expect to detect several IMBH mergers per year with upcoming GW observatories like LISA, LIGO's Voyager, the Einstein Telescope, and the Cosmic Explorer. Analysis of the merger events will provide important constraints on the underlying astrophysical populations of the merging binaries and will help constrain the IMBH formation and growth processes in dense star clusters.

\section*{acknowledgements}
    R.A.P.\ acknowledges support from the Student Experiential Learning Fund and the Undergraduate Advising and Research at Dartmouth College. This work was supported by NASA Grant 80NSSC21K1722 and NSF Grant AST-2108624 at Northwestern University (to G.F.\ and F.A.R.). We thank Miguel Angel Martinez, Quinn O. Casey, and Emmanuel Durodola for insightful discussions. R.A.P.~was a summer student in the CIERA Summer REU program directed by Dr.~Aaron Geller and supported by NSF Grant AST-2149425. 
    This research was supported in part through the computational resources and staff contributions provided for the Quest high performance computing facility at Northwestern University which is jointly supported by the Office of the Provost, the Office for Research, and Northwestern University Information Technology. 
    Much of this research was carried out at Evanston, IL, which is home to the Potawatomi, Odawa and Ojibwe Tribes, also known as the Niswimishkodewinan, an alliance of Anishinaabeg peoples.

\software{\code{scipy} \citep{2020SciPy-NMeth}, \code{astropy} \citep{Astropy2018}, \code{imbhistory} \citep{2023zndoFragione}}

\bibliographystyle{aasjournal}
\bibliography{bib}

\begin{thebibliography}{}
\expandafter\ifx\csname natexlab\endcsname\relax\def\natexlab#1{#1}\fi
\providecommand{\url}[1]{\href{#1}{#1}}
\providecommand{\dodoi}[1]{doi:~\href{http://doi.org/#1}{\nolinkurl{#1}}}
\providecommand{\doeprint}[1]{\href{http://ascl.net/#1}{\nolinkurl{http://ascl.net/#1}}}
\providecommand{\doarXiv}[1]{\href{https://arxiv.org/abs/#1}{\nolinkurl{https://arxiv.org/abs/#1}}}

\bibitem[{{Abbott} {et~al.}(2019){Abbott}, {Abbott}, {Abbott}, {Abraham},
  {Acernese}, {Ackley}, {Adams}, {Adams}, {Adhikari}, {Adya}, {Affeldt},
  {Agathos}, {Agatsuma}, {Aggarwal}, {Aguiar}, {Aiello}, {Ain}, {Ajith},
  {Allen}, {Allocca}, {Aloy}, {Altin}, {Amato}, {Anand}, {Ananyeva},
  {Anderson}, {Anderson}, {Angelova}, {Antier}, {Appert}, {Arai}, {Araya},
  {Areeda}, {Ar{\`e}ne}, {Arnaud}, {Aronson}, {Arun}, {Ascenzi}, {Ashton},
  {Aston}, {Astone}, {Aubin}, {Aufmuth}, {AultONeal}, {Austin}, {Avendano},
  {Avila-Alvarez}, {Babak}, {Bacon}, {Badaracco}, {Bader}, {Bae}, {Baer},
  {Baird}, {Baker}, {Baldaccini}, {Ballardin}, {Ballmer}, {Bals}, {Banagiri},
  {Barayoga}, {Barbieri}, {Barclay}, {Barish}, {Barker}, {Barkett}, {Barnum},
  {Barone}, {Barr}, {Barsotti}, {Barsuglia}, {Barta}, {Bartlett}, {Bartos},
  {Bassiri}, {Basti}, {Bawaj}, {Bayley}, {Bazzan}, {B{\'e}csy}, {Bejger},
  {Belahcene}, {Bell}, {Beniwal}, {Benjamin}, {Berger}, {Bergmann}, {Bernuzzi},
  {Berry}, {Bersanetti}, {Bertolini}, {Betzwieser}, {Bhandare}, {Bidler},
  {Biggs}, {Bilenko}, {Bilgili}, {Billingsley}, {Birney}, {Birnholtz},
  {Biscans}, {Bischi}, {Biscoveanu}, {Bisht}, {Bitossi}, {Bizouard},
  {Blackburn}, {Blackman}, {Blair}, {Blair}, {Blair}, {Bloemen}, {Bobba},
  {Bode}, {Boer}, {Boetzel}, {Bogaert}, {Bondu}, {Bonnand}, {Booker}, {Boom},
  {Bork}, {Boschi}, {Bose}, {Bossilkov}, {Bosveld}, {Bouffanais}, {Bozzi},
  {Bradaschia}, {Brady}, {Bramley}, {Branchesi}, {Brau}, {Breschi}, {Briant},
  {Briggs}, {Brighenti}, {Brillet}, {Brinkmann}, {Brockill}, {Brooks},
  {Brooks}, {Brown}, {Brunett}, {Buikema}, {Bulik}, {Bulten}, {Buonanno},
  {Buskulic}, {Buy}, {Byer}, {Cabero}, {Cadonati}, {Cagnoli}, {Cahillane},
  {Calder{\'o}n Bustillo}, {Callister}, {Calloni}, {Camp}, {Campbell},
  {Canepa}, {Cannon}, {Cao}, {Cao}, {Carapella}, {Carbognani}, {Caride},
  {Carney}, {Carullo}, {Casanueva Diaz}, {Casentini}, {Caudill},
  {Cavagli{\`a}}, {Cavalier}, {Cavalieri}, {Cella}, {Cerd{\'a}-Dur{\'a}n},
  {Cesarini}, {Chaibi}, {Chakravarti}, {Chamberlin}, {Chan}, {Chao},
  {Charlton}, {Chase}, {Chassande-Mottin}, {Chatterjee}, {Chaturvedi},
  {Cheeseboro}, {Chen}, {Chen}, {Chen}, {Cheng}, {Cheong}, {Chia}, {Chiadini},
  {Chincarini}, {Chiummo}, {Cho}, {Cho}, {Cho}, {Christensen}, {Chu}, {Chua},
  {Chung}, {Chung}, {Ciani}, {Cie{\'s}lar}, {Ciobanu}, {Ciolfi}, {Cipriano},
  {Cirone}, {Clara}, {Clark}, {Clearwater}, {Cleva}, {Coccia}, {Cohadon},
  {Cohen}, {Colleoni}, {Collette}, {Collins}, {Colpi}, {Cominsky},
  {Constancio}, {Conti}, {Cooper}, {Corban}, {Corbitt}, {Cordero-Carri{\'o}n},
  {Corezzi}, {Corley}, {Cornish}, {Corre}, {Corsi}, {Cortese}, {Costa},
  {Cotesta}, {Coughlin}, {Coughlin}, {Coulon}, {Countryman}, {Couvares},
  {Covas}, {Cowan}, {Coward}, {Cowart}, {Coyne}, {Coyne}, {Creighton},
  {Creighton}, {Cripe}, {Croquette}, {Crowder}, {Cullen}, {Cumming},
  {Cunningham}, {Cuoco}, {Canton}, {D{\'a}lya}, {D'Angelo}, {Danilishin},
  {D'Antonio}, {Danzmann}, {Dasgupta}, {Da Silva Costa}, {Datrier}, {Dattilo},
  {Dave}, {Davier}, {Davis}, {Daw}, {DeBra}, {Deenadayalan}, {Degallaix}, {De
  Laurentis}, {Del{\'e}glise}, {Del Pozzo}, {DeMarchi}, {Demos}, {Dent}, {De
  Pietri}, {De Rosa}, {De Rossi}, {DeSalvo}, {de Varona}, {Dhurandhar},
  {D{\'\i}az}, {Dietrich}, {Di Fiore}, {DiFronzo}, {Di Giorgio}, {Di Giovanni},
  {Di Giovanni}, {Di Girolamo}, {Di Lieto}, {Ding}, {Di Pace}, {Di Palma}, {Di
  Renzo}, {Divakarla}, {Dmitriev}, {Doctor}, {Donovan}, {Dooley}, {Doravari},
  {Dorrington}, {Downes}, {Drago}, {Driggers}, {Du}, {Ducoin}, {Dupej},
  {Durante}, {Dwyer}, {Easter}, {Eddolls}, {Edo}, {Effler}, {Ehrens},
  {Eichholz}, {Eikenberry}, {Eisenmann}, {Eisenstein}, {Errico}, {Essick},
  {Estelles}, {Estevez}, {Etienne}, {Etzel}, {Evans}, {Evans}, {Fafone},
  {Fairhurst}, {Fan}, {Farinon}, {Farr}, {Farr}, {Fauchon-Jones}, {Favata},
  {Fays}, {Fazio}, {Fee}, {Feicht}, {Fejer}, {Feng}, {Ferguson},
  {Fernandez-Galiana}, {Ferrante}, {Ferreira}, {Ferreira}, {Fidecaro}, {Fiori},
  {Fiorucci}, {Fishbach}, {Fisher}, {Fishner}, {Fittipaldi}, {Fitz-Axen},
  {Fiumara}, {Flaminio}, {Fletcher}, {Floden}, {Flynn}, {Fong}, {Font},
  {Forsyth}, {Fournier}, {Vivanco}, {Frasca}, {Frasconi}, {Frei}, {Freise},
  {Frey}, {Frey}, {Fritschel}, {Frolov}, {Fronz{\`e}}, {Fulda}, {Fyffe},
  {Gabbard}, {Gadre}, {Gaebel}, {Gair}, {Gammaitoni}, {Gaonkar},
  {Garc{\'\i}a-Quir{\'o}s}, {Garufi}, {Gateley}, {Gaudio}, {Gaur}, {Gayathri},
  {Gemme}, {Genin}, {Gennai}, {George}, {George}, {Gergely}, {Ghonge}, {Ghosh},
  {Ghosh}, {Ghosh}, {Giacomazzo}, {Giaime}, {Giardina}, {Gibson}, {Gill},
  {Glover}, {Gniesmer}, {Godwin}, {Goetz}, {Goetz}, {Goncharov},
  {Gonz{\'a}lez}, {Gonzalez Castro}, {Gopakumar}, {Gossan}, {Gosselin},
  {Gouaty}, {Grace}, {Grado}, {Granata}, {Grant}, {Gras}, {Grassia}, {Gray},
  {Gray}, {Greco}, {Green}, {Green}, {Gretarsson}, {Grimaldi}, {Grimm},
  {Groot}, {Grote}, {Grunewald}, {Gruning}, {Guidi}, {Gulati}, {Guo}, {Gupta},
  {Gupta}, {Gupta}, {Gustafson}, {Gustafson}, {Haegel}, {Halim}, {Hall},
  {Hall}, {Hamilton}, {Hammond}, {Haney}, {Hanke}, {Hanks}, {Hanna}, {Hannam},
  {Hannuksela}, {Hansen}, {Hanson}, {Harder}, {Hardwick}, {Haris}, {Harms},
  {Harry}, {Harry}, {Hasskew}, {Haster}, {Haughian}, {Hayes}, {Healy},
  {Heidmann}, {Heintze}, {Heitmann}, {Hellman}, {Hello}, {Hemming}, {Hendry},
  {Heng}, {Hennig}, {Heurs}, {Hild}, {Hinderer}, {Hochheim}, {Hofman},
  {Holgado}, {Holland}, {Holt}, {Holz}, {Hopkins}, {Horst}, {Hough}, {Howell},
  {Hoy}, {Huang}, {H{\"u}bner}, {Huerta}, {Huet}, {Hughey}, {Hui}, {Husa},
  {Huttner}, {Huynh-Dinh}, {Idzkowski}, {Iess}, {Inchauspe}, {Ingram}, {Inta},
  {Intini}, {Irwin}, {Isa}, {Isac}, {Isi}, {Iyer}, {Jacqmin}, {Jadhav}, {Jani},
  {Janthalur}, {Jaranowski}, {Jariwala}, {Jenkins}, {Jiang}, {Johns},
  {Johnson}, {Jones}, {Jones}, {Jones}, {Jones}, {Jonker}, {Ju}, {Junker},
  {Kalaghatgi}, {Kalogera}, {Kamai}, {Kandhasamy}, {Kang}, {Kanner}, {Kapadia},
  {Karki}, {Kashyap}, {Kasprzack}, {Katsanevas}, {Katsavounidis}, {Katzman},
  {Kaufer}, {Kawabe}, {Keerthana}, {K{\'e}f{\'e}lian}, {Keitel}, {Kennedy},
  {Key}, {Khalili}, {Khamesra}, {Khan}, {Khan}, {Khazanov}, {Khetan},
  {Khursheed}, {Kijbunchoo}, {Kim}, {Kim}, {Kim}, {Kim}, {Kim}, {Kim}, {Kim},
  {Kimball}, {King}, {Kinley-Hanlon}, {Kirchhoff}, {Kissel}, {Kleybolte},
  {Klika}, {Klimenko}, {Knowles}, {Koch}, {Koehlenbeck}, {Koekoek}, {Koley},
  {Kondrashov}, {Kontos}, {Koper}, {Korobko}, {Korth}, {Kovalam}, {Kozak},
  {Kr{\"a}mer}, {Kringel}, {Krishnendu}, {Kr{\'o}lak}, {Krupinski}, {Kuehn},
  {Kumar}, {Kumar}, {Kumar}, {Kumar}, {Kuo}, {Kutynia}, {Kwang}, {Lackey},
  {Laghi}, {Laguna}, {Lai}, {Lam}, {Landry}, {Lane}, {Lang}, {Lange}, {Lantz},
  {Lanza}, {Lartaux-Vollard}, {Lasky}, {Laxen}, {Lazzarini}, {Lazzaro},
  {Leaci}, {Leavey}, {Lecoeuche}, {Lee}, {Lee}, {Lee}, {Lee}, {Lee}, {Lee},
  {Lehmann}, {Lenon}, {Leroy}, {Letendre}, {Levin}, {Li}, {Li}, {Li}, {Li},
  {Li}, {Lin}, {Linde}, {Linker}, {Littenberg}, {Liu}, {Liu},
  {Llorens-Monteagudo}, {Lo}, {London}, {Longo}, {Lorenzini}, {Loriette},
  {Lormand}, {Losurdo}, {Lough}, {Lousto}, {Lovelace}, {Lower}, {L{\"u}ck},
  {Lumaca}, {Lundgren}, {Lynch}, {Ma}, {Macas}, {Macfoy}, {MacInnis},
  {Macleod}, {Macquet}, {Maga{\~n}a Hernandez}, {Maga{\~n}a-Sandoval}, {Magee},
  {Majorana}, {Maksimovic}, {Malik}, {Man}, {Mandic}, {Mangano}, {Mansell},
  {Manske}, {Mantovani}, {Mapelli}, {Marchesoni}, {Marion}, {M{\'a}rka},
  {M{\'a}rka}, {Markakis}, {Markosyan}, {Markowitz}, {Maros}, {Marquina},
  {Marsat}, {Martelli}, {Martin}, {Martin}, {Martinez}, {Martynov},
  {Masalehdan}, {Mason}, {Massera}, {Masserot}, {Massinger}, {Masso-Reid},
  {Mastrogiovanni}, {Matas}, {Matichard}, {Matone}, {Mavalvala}, {McCann},
  {McCarthy}, {McClelland}, {McCormick}, {McCuller}, {McGuire}, {McIsaac},
  {McIver}, {McManus}, {McRae}, {McWilliams}, {Meacher}, {Meadors}, {Mehmet},
  {Mehta}, {Meidam}, {Mejuto Villa}, {Melatos}, {Mendell}, {Mercer}, {Mereni},
  {Merfeld}, {Merilh}, {Merzougui}, {Meshkov}, {Messenger}, {Messick},
  {Messina}, {Metzdorff}, {Meyers}, {Meylahn}, {Miani}, {Miao}, {Michel},
  {Middleton}, {Milano}, {Miller}, {Millhouse}, {Mills}, {Milovich-Goff},
  {Minazzoli}, {Minenkov}, {Mishkin}, {Mishra}, {Mistry}, {Mitra},
  {Mitrofanov}, {Mitselmakher}, {Mittleman}, {Mo}, {Moffa}, {Mogushi},
  {Mohapatra}, {Molina-Ruiz}, {Mondin}, {Montani}, {Moore}, {Moraru},
  {Morawski}, {Moreno}, {Morisaki}, {Mours}, {Mow-Lowry}, {Muciaccia},
  {Mukherjee}, {Mukherjee}, {Mukherjee}, {Mukherjee}, {Mukund}, {Mullavey},
  {Munch}, {Mu{\~n}iz}, {Muratore}, {Murray}, {Nagar}, {Nardecchia},
  {Naticchioni}, {Nayak}, {Neil}, {Neilson}, {Nelemans}, {Nelson}, {Nery},
  {Neunzert}, {Nevin}, {Ng}, {Ng}, {Nguyen}, {Nguyen}, {Nichols}, {Nichols},
  {Nissanke}, {Nocera}, {North}, {Nuttall}, {Obergaulinger}, {Oberling},
  {O'Brien}, {Oganesyan}, {Ogin}, {Oh}, {Oh}, {Ohme}, {Ohta}, {Okada},
  {Oliver}, {Oppermann}, {Oram}, {O'Reilly}, {Ormiston}, {Ortega},
  {O'Shaughnessy}, {Ossokine}, {Ottaway}, {Overmier}, {Owen}, {Pace}, {Pagano},
  {Page}, {Pagliaroli}, {Pai}, {Pai}, {Palamos}, {Palashov}, {Palomba}, {Pan},
  {Panda}, {Pang}, {Pankow}, {Pannarale}, {Pant}, {Paoletti}, {Paoli},
  {Parida}, {Parker}, {Pascucci}, {Pasqualetti}, {Passaquieti}, {Passuello},
  {Patil}, {Patricelli}, {Payne}, {Pearlstone}, {Pechsiri}, {Pedersen},
  {Pedraza}, {Pedurand}, {Pele}, {Penn}, {Perego}, {Perez}, {P{\'e}rigois},
  {Perreca}, {Petermann}, {Pfeiffer}, {Phelps}, {Phukon}, {Piccinni}, {Pichot},
  {Piergiovanni}, {Pierro}, {Pillant}, {Pinard}, {Pinto}, {Pirello}, {Pitkin},
  {Plastino}, {Poggiani}, {Pong}, {Ponrathnam}, {Popolizio}, {Porter},
  {Powell}, {Prajapati}, {Prasad}, {Prasai}, {Prasanna}, {Pratten},
  {Prestegard}, {Principe}, {Prodi}, {Prokhorov}, {Punturo}, {Puppo},
  {P{\"u}rrer}, {Qi}, {Quetschke}, {Quinonez}, {Raab}, {Raaijmakers},
  {Radkins}, {Radulesco}, {Raffai}, {Raja}, {Rajan}, {Rajbhandari},
  {Rakhmanov}, {Ramirez}, {Ramos-Buades}, {Rana}, {Rao}, {Rapagnani},
  {Raymond}, {Razzano}, {Read}, {Regimbau}, {Rei}, {Reid}, {Reitze},
  {Rettegno}, {Ricci}, {Richardson}, {Richardson}, {Ricker}, {Riemenschneider},
  {Riles}, {Rizzo}, {Robertson}, {Robinet}, {Rocchi}, {Rolland}, {Rollins},
  {Roma}, {Romanelli}, {Romano}, {Romel}, {Romie}, {Rose}, {Rose}, {Rose},
  {Rosi{\'n}ska}, {Rosofsky}, {Ross}, {Rowan}, {R{\"u}diger}, {Ruggi},
  {Rutins}, {Ryan}, {Sachdev}, {Sadecki}, {Sakellariadou}, {Salafia},
  {Salconi}, {Saleem}, {Samajdar}, {Sammut}, {Sanchez}, {Sanchez},
  {Sanchis-Gual}, {Sanders}, {Santiago}, {Santos}, {Sarin}, {Sassolas},
  {Sathyaprakash}, {Sauter}, {Savage}, {Schale}, {Scheel}, {Scheuer},
  {Schmidt}, {Schnabel}, {Schofield}, {Sch{\"o}nbeck}, {Schreiber}, {Schulte},
  {Schutz}, {Scott}, {Scott}, {Seidel}, {Sellers}, {Sengupta}, {Sennett},
  {Sentenac}, {Sequino}, {Sergeev}, {Setyawati}, {Shaddock}, {Shaffer},
  {Shahriar}, {Shaner}, {Sharma}, {Sharma}, {Shawhan}, {Shen}, {Shink},
  {Shoemaker}, {Shoemaker}, {Shukla}, {ShyamSundar}, {Siellez}, {Sieniawska},
  {Sigg}, {Singer}, {Singh}, {Singh}, {Singhal}, {Sintes}, {Sitmukhambetov},
  {Skliris}, {Slagmolen}, {Slaven-Blair}, {Smith}, {Smith}, {Somala}, {Son},
  {Soni}, {Sorazu}, {Sorrentino}, {Souradeep}, {Sowell}, {Spencer}, {Spera},
  {Srivastava}, {Srivastava}, {Staats}, {Stachie}, {Standke}, {Steer},
  {Steinke}, {Steinlechner}, {Steinlechner}, {Steinmeyer}, {Stevenson},
  {Stocks}, {Stolle-McAllister}, {Stone}, {Stops}, {Strain}, {Stratta},
  {Strigin}, {Strunk}, {Sturani}, {Stuver}, {Sudhir}, {Summerscales}, {Sun},
  {Sunil}, {Sur}, {Suresh}, {Sutton}, {Swinkels}, {Szczepa{\'n}czyk}, {Tacca},
  {Tait}, {Talbot}, {Tanner}, {Tao}, {T{\'a}pai}, {Tapia}, {Tasson}, {Taylor},
  {Tenorio}, {Terkowski}, {Thomas}, {Thomas}, {Thondapu}, {Thorne}, {Thrane},
  {Tiwari}, {Tiwari}, {Tiwari}, {Toland}, {Tonelli}, {Tornasi},
  {Torres-Forn{\'e}}, {Torrie}, {T{\"o}yr{\"a}}, {Travasso}, {Traylor},
  {Tringali}, {Tripathee}, {Trovato}, {Trozzo}, {Tsang}, {Tse}, {Tso},
  {Tsukada}, {Tsuna}, {Tsutsui}, {Tuyenbayev}, {Ueno}, {Ugolini},
  {Unnikrishnan}, {Urban}, {Usman}, {Vahlbruch}, {Vajente}, {Valdes},
  {Valentini}, {van Bakel}, {van Beuzekom}, {van den Brand}, {Van Den Broeck},
  {Vander-Hyde}, {van der Schaaf}, {VanHeijningen}, {van Veggel}, {Vardaro},
  {Varma}, {Vass}, {Vas{\'u}th}, {Vecchio}, {Vedovato}, {Veitch}, {Veitch},
  {Venkateswara}, {Venugopalan}, {Verkindt}, {Vetrano}, {Vicer{\'e}}, {Viets},
  {Vinciguerra}, {Vine}, {Vinet}, {Vitale}, {Vo}, {Vocca}, {Vorvick},
  {Vyatchanin}, {Wade}, {Wade}, {Wade}, {Walet}, {Walker}, {Wallace}, {Walsh},
  {Wang}, {Wang}, {Wang}, {Wang}, {Wang}, {Ward}, {Warden}, {Warner}, {Was},
  {Watchi}, {Weaver}, {Wei}, {Weinert}, {Weinstein}, {Weiss}, {Wellmann},
  {Wen}, {Wessel}, {We{\ss}els}, {Westhouse}, {Wette}, {Whelan}, {Whiting},
  {Whittle}, {Wilken}, {Williams}, {Williamson}, {Willis}, {Willke}, {Winkler},
  {Wipf}, {Wittel}, {Woan}, {Woehler}, {Wofford}, {Wright}, {Wu}, {Wysocki},
  {Xiao}, {Xu}, {Yamamoto}, {Yancey}, {Yang}, {Yang}, {Yang}, {Yap}, {Yazback},
  {Yeeles}, {Yoon}, {Yu}, {Yu}, {Yuen}, {Zadro{\.Z}ny}, {Zadro{\.Z}ny},
  {Zanolin}, {Zelenova}, {Zendri}, {Zevin}, {Zhang}, {Zhang}, {Zhang}, {Zhao},
  {Zhao}, {Zhou}, {Zhou}, {Zhu}, {Zucker}, {Zweizig}, {Salemi}, {Papa}, {LIGO
  Scientific Collaboration}, \& {Virgo Collaboration}}]{Abbott2019}
{Abbott}, B.~P., {Abbott}, R., {Abbott}, T.~D., {et~al.} 2019, \prd, 100,
  064064, \dodoi{10.1103/PhysRevD.100.064064}

\bibitem[{{Abbott} {et~al.}(2020){Abbott}, {Abbott}, {Abraham}, {Acernese},
  {Ackley}, {Adams}, {Adhikari}, {Adya}, {Affeldt}, {Agathos}, {Agatsuma},
  {Aggarwal}, {Aguiar}, {Aich}, {Aiello}, {Ain}, {Ajith}, {Akcay}, {Allen},
  {Allocca}, {Altin}, {Amato}, {Anand}, {Ananyeva}, {Anderson}, {Anderson},
  {Angelova}, {Ansoldi}, {Antier}, {Appert}, {Arai}, {Araya}, {Areeda},
  {Ar{\`e}ne}, {Arnaud}, {Aronson}, {Arun}, {Asali}, {Ascenzi}, {Ashton},
  {Aston}, {Astone}, {Aubin}, {Aufmuth}, {AultONeal}, {Austin}, {Avendano},
  {Babak}, {Bacon}, {Badaracco}, {Bader}, {Bae}, {Baer}, {Baird}, {Baldaccini},
  {Ballardin}, {Ballmer}, {Bals}, {Balsamo}, {Baltus}, {Banagiri}, {Bankar},
  {Bankar}, {Barayoga}, {Barbieri}, {Barish}, {Barker}, {Barkett}, {Barneo},
  {Barone}, {Barr}, {Barsotti}, {Barsuglia}, {Barta}, {Bartlett}, {Bartos},
  {Bassiri}, {Basti}, {Bawaj}, {Bayley}, {Bazzan}, {B{\'e}csy}, {Bejger},
  {Belahcene}, {Bell}, {Beniwal}, {Benjamin}, {Bentley}, {Bergamin}, {Berger},
  {Bergmann}, {Bernuzzi}, {Berry}, {Bersanetti}, {Bertolini}, {Betzwieser},
  {Bhandare}, {Bhandari}, {Bidler}, {Biggs}, {Bilenko}, {Billingsley},
  {Birney}, {Birnholtz}, {Biscans}, {Bischi}, {Biscoveanu}, {Bisht},
  {Bissenbayeva}, {Bitossi}, {Bizouard}, {Blackburn}, {Blackman}, {Blair},
  {Blair}, {Blair}, {Bobba}, {Bode}, {Boer}, {Boetzel}, {Bogaert}, {Bondu},
  {Bonilla}, {Bonnand}, {Booker}, {Boom}, {Bork}, {Boschi}, {Bose},
  {Bossilkov}, {Bosveld}, {Bouffanais}, {Bozzi}, {Bradaschia}, {Brady},
  {Bramley}, {Branchesi}, {Brau}, {Breschi}, {Briant}, {Briggs}, {Brighenti},
  {Brillet}, {Brinkmann}, {Brockill}, {Brooks}, {Brooks}, {Brown}, {Brunett},
  {Bruno}, {Bruntz}, {Buikema}, {Bulik}, {Bulten}, {Buonanno}, {Buscicchio},
  {Buskulic}, {Byer}, {Cabero}, {Cadonati}, {Cagnoli}, {Cahillane},
  {Calder{\'o}n Bustillo}, {Callaghan}, {Callister}, {Calloni}, {Camp},
  {Canepa}, {Cannon}, {Cao}, {Cao}, {Carapella}, {Carbognani}, {Caride},
  {Carney}, {Carullo}, {Casanueva Diaz}, {Casentini}, {Casta{\~n}eda},
  {Caudill}, {Cavagli{\`a}}, {Cavalier}, {Cavalieri}, {Cella},
  {Cerd{\'a}-Dur{\'a}n}, {Cesarini}, {Chaibi}, {Chakravarti}, {Chan}, {Chan},
  {Chandra}, {Chao}, {Charlton}, {Chase}, {Chassande-Mottin}, {Chatterjee},
  {Chaturvedi}, {Chatziioannou}, {Chen}, {Chen}, {Chen}, {Cheng}, {Cheong},
  {Chia}, {Chiadini}, {Chierici}, {Chincarini}, {Chiummo}, {Cho}, {Cho}, {Cho},
  {Christensen}, {Chu}, {Chua}, {Chung}, {Chung}, {Ciani}, {Ciecielag},
  {Cie{\'s}lar}, {Ciobanu}, {Ciolfi}, {Cipriano}, {Cirone}, {Clara}, {Clark},
  {Clearwater}, {Clesse}, {Cleva}, {Coccia}, {Cohadon}, {Cohen}, {Colleoni},
  {Collette}, {Collins}, {Colpi}, {Constancio}, {Conti}, {Cooper}, {Corban},
  {Corbitt}, {Cordero-Carri{\'o}n}, {Corezzi}, {Corley}, {Cornish}, {Corre},
  {Corsi}, {Cortese}, {Costa}, {Cotesta}, {Coughlin}, {Coughlin}, {Coulon},
  {Countryman}, {Couvares}, {Covas}, {Coward}, {Cowart}, {Coyne}, {Coyne},
  {Creighton}, {Creighton}, {Cripe}, {Croquette}, {Crowder}, {Cudell},
  {Cullen}, {Cumming}, {Cummings}, {Cunningham}, {Cuoco}, {Curylo}, {Canton},
  {D{\'a}lya}, {Dana}, {Daneshgaran-Bajastani}, {D'Angelo}, {Danilishin},
  {D'Antonio}, {Danzmann}, {Darsow-Fromm}, {Dasgupta}, {Datrier}, {Dattilo},
  {Dave}, {Davier}, {Davies}, {Davis}, {Daw}, {DeBra}, {Deenadayalan},
  {Degallaix}, {De Laurentis}, {Del{\'e}glise}, {Delfavero}, {De Lillo}, {Del
  Pozzo}, {DeMarchi}, {D'Emilio}, {Demos}, {Dent}, {De Pietri}, {De Rosa}, {De
  Rossi}, {DeSalvo}, {de Varona}, {Dhurandhar}, {D{\'\i}az}, {Diaz-Ortiz},
  {Dietrich}, {Di Fiore}, {Di Fronzo}, {Di Giorgio}, {Di Giovanni}, {Di
  Giovanni}, {Di Girolamo}, {Di Lieto}, {Ding}, {Di Pace}, {Di Palma}, {Di
  Renzo}, {Divakarla}, {Dmitriev}, {Doctor}, {Donovan}, {Dooley}, {Doravari},
  {Dorrington}, {Downes}, {Drago}, {Driggers}, {Du}, {Ducoin}, {Dupej},
  {Durante}, {D'Urso}, {Dwyer}, {Easter}, {Eddolls}, {Edelman}, {Edo}, {Edy},
  {Effler}, {Ehrens}, {Eichholz}, {Eikenberry}, {Eisenmann}, {Eisenstein},
  {Ejlli}, {Errico}, {Essick}, {Estelles}, {Estevez}, {Etienne}, {Etzel},
  {Evans}, {Evans}, {Ewing}, {Fafone}, {Fairhurst}, {Fan}, {Farinon}, {Farr},
  {Farr}, {Fauchon-Jones}, {Favata}, {Fays}, {Fazio}, {Feicht}, {Fejer},
  {Feng}, {Fenyvesi}, {Ferguson}, {Fernandez-Galiana}, {Ferrante}, {Ferreira},
  {Ferreira}, {Fidecaro}, {Fiori}, {Fiorucci}, {Fishbach}, {Fisher},
  {Fittipaldi}, {Fitz-Axen}, {Fiumara}, {Flaminio}, {Floden}, {Flynn}, {Fong},
  {Font}, {Forsyth}, {Fournier}, {Frasca}, {Frasconi}, {Frei}, {Freise},
  {Frey}, {Frey}, {Fritschel}, {Frolov}, {Fronz{\`e}}, {Fulda}, {Fyffe},
  {Gabbard}, {Gadre}, {Gaebel}, {Gair}, {Galaudage}, {Ganapathy}, {Ganguly},
  {Gaonkar}, {Garc{\'\i}a-Quir{\'o}s}, {Garufi}, {Gateley}, {Gaudio},
  {Gayathri}, {Gemme}, {Genin}, {Gennai}, {George}, {George}, {Gergely},
  {Ghonge}, {Ghosh}, {Ghosh}, {Ghosh}, {Giacomazzo}, {Giaime}, {Giardina},
  {Gibson}, {Gier}, {Gill}, {Glanzer}, {Gniesmer}, {Godwin}, {Goetz}, {Goetz},
  {Gohlke}, {Goncharov}, {Gonz{\'a}lez}, {Gopakumar}, {Gossan}, {Gosselin},
  {Gouaty}, {Grace}, {Grado}, {Granata}, {Grant}, {Gras}, {Grassia}, {Gray},
  {Gray}, {Greco}, {Green}, {Green}, {Gretarsson}, {Griggs}, {Grignani},
  {Grimaldi}, {Grimm}, {Grote}, {Grunewald}, {Gruning}, {Guidi}, {Guimaraes},
  {Guix{\'e}}, {Gulati}, {Guo}, {Gupta}, {Gupta}, {Gupta}, {Gustafson},
  {Gustafson}, {Haegel}, {Halim}, {Hall}, {Hamilton}, {Hammond}, {Haney},
  {Hanke}, {Hanks}, {Hanna}, {Hannam}, {Hannuksela}, {Hansen}, {Hanson},
  {Harder}, {Hardwick}, {Haris}, {Harms}, {Harry}, {Harry}, {Hasskew},
  {Haster}, {Haughian}, {Hayes}, {Healy}, {Heidmann}, {Heintze}, {Heinze},
  {Heitmann}, {Hellman}, {Hello}, {Hemming}, {Hendry}, {Heng}, {Hennes},
  {Hennig}, {Heurs}, {Hild}, {Hinderer}, {Hoback}, {Hochheim}, {Hofgard},
  {Hofman}, {Holgado}, {Holland}, {Holt}, {Holz}, {Hopkins}, {Horst}, {Hough},
  {Howell}, {Hoy}, {Huang}, {H{\"u}bner}, {Huerta}, {Huet}, {Hughey}, {Hui},
  {Husa}, {Huttner}, {Huxford}, {Huynh-Dinh}, {Idzkowski}, {Iess}, {Inchauspe},
  {Ingram}, {Intini}, {Isac}, {Isi}, {Iyer}, {Jacqmin}, {Jadhav}, {Jadhav},
  {James}, {Jani}, {Janthalur}, {Jaranowski}, {Jariwala}, {Jaume}, {Jenkins},
  {Jiang}, {Johns}, {Johnson-McDaniel}, {Jones}, {Jones}, {Jones}, {Jones},
  {Jones}, {Jonker}, {Ju}, {Junker}, {Kalaghatgi}, {Kalogera}, {Kamai},
  {Kandhasamy}, {Kang}, {Kanner}, {Kapadia}, {Karki}, {Kashyap}, {Kasprzack},
  {Kastaun}, {Katsanevas}, {Katsavounidis}, {Katzman}, {Kaufer}, {Kawabe},
  {K{\'e}f{\'e}lian}, {Keitel}, {Keivani}, {Kennedy}, {Key}, {Khadka},
  {Khalili}, {Khan}, {Khan}, {Khan}, {Khazanov}, {Khetan}, {Khursheed},
  {Kijbunchoo}, {Kim}, {Kim}, {Kim}, {Kim}, {Kim}, {Kim}, {Kim}, {Kimball},
  {King}, {Kinley-Hanlon}, {Kirchhoff}, {Kissel}, {Kleybolte}, {Klimenko},
  {Knowles}, {Knyazev}, {Koch}, {Koehlenbeck}, {Koekoek}, {Koley},
  {Kondrashov}, {Kontos}, {Koper}, {Korobko}, {Korth}, {Kovalam}, {Kozak},
  {Kringel}, {Krishnendu}, {Kr{\'o}lak}, {Krupinski}, {Kuehn}, {Kumar},
  {Kumar}, {Kumar}, {Kumar}, {Kumar}, {Kuo}, {Kutynia}, {Lackey}, {Laghi},
  {Lalande}, {Lam}, {Lamberts}, {Landry}, {Lane}, {Lang}, {Lange}, {Lantz},
  {Lanza}, {La Rosa}, {Lartaux-Vollard}, {Lasky}, {Laxen}, {Lazzarini},
  {Lazzaro}, {Leaci}, {Leavey}, {Lecoeuche}, {Lee}, {Lee}, {Lee}, {Lee}, {Lee},
  {Lehmann}, {Leroy}, {Letendre}, {Levin}, {Li}, {Li}, {li}, {Li}, {Li},
  {Linde}, {Linker}, {Linley}, {Littenberg}, {Liu}, {Liu},
  {Llorens-Monteagudo}, {Lo}, {Lockwood}, {London}, {Longo}, {Lorenzini},
  {Loriette}, {Lormand}, {Losurdo}, {Lough}, {Lousto}, {Lovelace}, {L{\"u}ck},
  {Lumaca}, {Lundgren}, {Ma}, {Macas}, {Macfoy}, {MacInnis}, {Macleod},
  {MacMillan}, {Macquet}, {Maga{\~n}a Hernandez}, {Maga{\~n}a-Sandoval},
  {Magee}, {Majorana}, {Maksimovic}, {Malik}, {Man}, {Mandic}, {Mangano},
  {Mansell}, {Manske}, {Mantovani}, {Mapelli}, {Marchesoni}, {Marion},
  {M{\'a}rka}, {M{\'a}rka}, {Markakis}, {Markosyan}, {Markowitz}, {Maros},
  {Marquina}, {Marsat}, {Martelli}, {Martin}, {Martin}, {Martinez}, {Martynov},
  {Masalehdan}, {Mason}, {Massera}, {Masserot}, {Massinger}, {Masso-Reid},
  {Mastrogiovanni}, {Matas}, {Matichard}, {Mavalvala}, {Maynard}, {McCann},
  {McCarthy}, {McClelland}, {McCormick}, {McCuller}, {McGuire}, {McIsaac},
  {McIver}, {McManus}, {McRae}, {McWilliams}, {Meacher}, {Meadors}, {Mehmet},
  {Mehta}, {Mejuto Villa}, {Melatos}, {Mendell}, {Mercer}, {Mereni}, {Merfeld},
  {Merilh}, {Merritt}, {Merzougui}, {Meshkov}, {Messenger}, {Messick},
  {Metzdorff}, {Meyers}, {Meylahn}, {Mhaske}, {Miani}, {Miao}, {Michaloliakos},
  {Michel}, {Middleton}, {Milano}, {Miller}, {Millhouse}, {Mills}, {Milotti},
  {Milovich-Goff}, {Minazzoli}, {Minenkov}, {Mishkin}, {Mishra}, {Mistry},
  {Mitra}, {Mitrofanov}, {Mitselmakher}, {Mittleman}, {Mo}, {Mogushi},
  {Mohapatra}, {Mohite}, {Molina-Ruiz}, {Mondin}, {Montani}, {Moore}, {Moraru},
  {Morawski}, {Moreno}, {Morisaki}, {Mours}, {Mow-Lowry}, {Mozzon},
  {Muciaccia}, {Mukherjee}, {Mukherjee}, {Mukherjee}, {Mukherjee}, {Mukund},
  {Mullavey}, {Munch}, {Mu{\~n}iz}, {Murray}, {Nagar}, {Nardecchia},
  {Naticchioni}, {Nayak}, {Neil}, {Neilson}, {Nelemans}, {Nelson}, {Nery},
  {Neunzert}, {Ng}, {Ng}, {Nguyen}, {Nguyen}, {Nichols}, {Nichols}, {Nissanke},
  {Nitz}, {Nocera}, {Noh}, {North}, {Nothard}, {Nuttall}, {Oberling},
  {O'Brien}, {Oganesyan}, {Ogin}, {Oh}, {Oh}, {Ohme}, {Ohta}, {Okada},
  {Oliver}, {Olivetto}, {Oppermann}, {Oram}, {O'Reilly}, {Ormiston}, {Ortega},
  {O'Shaughnessy}, {Ossokine}, {Osthelder}, {Ottaway}, {Overmier}, {Owen},
  {Pace}, {Pagano}, {Page}, {Pagliaroli}, {Pai}, {Pai}, {Palamos}, {Palashov},
  {Palomba}, {Pan}, {Panda}, {Pang}, {Pankow}, {Pannarale}, {Pant}, {Paoletti},
  {Paoli}, {Parida}, {Parker}, {Pascucci}, {Pasqualetti}, {Passaquieti},
  {Passuello}, {Patricelli}, {Payne}, {Pearlstone}, {Pechsiri}, {Pedersen},
  {Pedraza}, {Pele}, {Penn}, {Perego}, {Perez}, {P{\'e}rigois}, {Perreca},
  {Perri{\`e}s}, {Petermann}, {Pfeiffer}, {Phelps}, {Phukon}, {Piccinni},
  {Pichot}, {Piendibene}, {Piergiovanni}, {Pierro}, {Pillant}, {Pinard},
  {Pinto}, {Piotrzkowski}, {Pirello}, {Pitkin}, {Plastino}, {Poggiani}, {Pong},
  {Ponrathnam}, {Popolizio}, {Porter}, {Powell}, {Prajapati}, {Prasai},
  {Prasanna}, {Pratten}, {Prestegard}, {Principe}, {Prodi}, {Prokhorov},
  {Punturo}, {Puppo}, {P{\"u}rrer}, {Qi}, {Quetschke}, {Quinonez}, {Raab},
  {Raaijmakers}, {Radkins}, {Radulesco}, {Raffai}, {Rafferty}, {Raja}, {Rajan},
  {Rajbhandari}, {Rakhmanov}, {Ramirez}, {Ramos-Buades}, {Rana}, {Rao},
  {Rapagnani}, {Raymond}, {Razzano}, {Read}, {Regimbau}, {Rei}, {Reid},
  {Reitze}, {Rettegno}, {Ricci}, {Richardson}, {Richardson}, {Ricker},
  {Riemenschneider}, {Riles}, {Rizzo}, {Robertson}, {Robinet}, {Rocchi},
  {Rodriguez-Soto}, {Rolland}, {Rollins}, {Roma}, {Romanelli}, {Romano},
  {Romel}, {Romero-Shaw}, {Romie}, {Rose}, {Rose}, {Rose}, {Rosi{\'n}ska},
  {Rosofsky}, {Ross}, {Rowan}, {Rowlinson}, {Roy}, {Roy}, {Roy}, {Ruggi},
  {Rutins}, {Ryan}, {Sachdev}, {Sadecki}, {Sakellariadou}, {Salafia},
  {Salconi}, {Saleem}, {Salemi}, {Samajdar}, {Sanchez}, {Sanchez},
  {Sanchis-Gual}, {Sanders}, {Santiago}, {Santos}, {Sarin}, {Sassolas},
  {Sathyaprakash}, {Sauter}, {Savage}, {Savant}, {Sawant}, {Sayah}, {Schaetzl},
  {Schale}, {Scheel}, {Scheuer}, {Schmidt}, {Schnabel}, {Schofield},
  {Sch{\"o}nbeck}, {Schreiber}, {Schulte}, {Schutz}, {Schwarm}, {Schwartz},
  {Scott}, {Scott}, {Seidel}, {Sellers}, {Sengupta}, {Sennett}, {Sentenac},
  {Sequino}, {Sergeev}, {Setyawati}, {Shaddock}, {Shaffer}, {Sharifi},
  {Shahriar}, {Sharma}, {Sharma}, {Shawhan}, {Shen}, {Shikauchi}, {Shink},
  {Shoemaker}, {Shoemaker}, {Shukla}, {ShyamSundar}, {Siellez}, {Sieniawska},
  {Sigg}, {Singer}, {Singh}, {Singh}, {Singha}, {Singhal}, {Sintes}, {Sipala},
  {Skliris}, {Slagmolen}, {Slaven-Blair}, {Smetana}, {Smith}, {Smith},
  {Somala}, {Son}, {Soni}, {Sorazu}, {Sordini}, {Sorrentino}, {Souradeep},
  {Sowell}, {Spencer}, {Spera}, {Srivastava}, {Srivastava}, {Staats},
  {Stachie}, {Standke}, {Steer}, {Steinke}, {Steinlechner}, {Steinlechner},
  {Steinmeyer}, {Stevenson}, {Stocks}, {Stops}, {Stover}, {Strain}, {Stratta},
  {Strunk}, {Sturani}, {Stuver}, {Sudhagar}, {Sudhir}, {Summerscales}, {Sun},
  {Sunil}, {Sur}, {Suresh}, {Sutton}, {Swinkels}, {Szczepa{\'n}czyk}, {Tacca},
  {Tait}, {Talbot}, {Tanasijczuk}, {Tanner}, {Tao}, {T{\'a}pai}, {Tapia},
  {Tapia San Martin}, {Tasson}, {Taylor}, {Tenorio}, {Terkowski},
  {Thirugnanasambandam}, {Thomas}, {Thomas}, {Thompson}, {Thondapu}, {Thorne},
  {Thrane}, {Tinsman}, {Saravanan}, {Tiwari}, {Tiwari}, {Tiwari}, {Toland},
  {Tonelli}, {Tornasi}, {Torres-Forn{\'e}}, {Torrie}, {Tosta e Melo},
  {T{\"o}yr{\"a}}, {Travasso}, {Traylor}, {Tringali}, {Tripathee}, {Trovato},
  {Trudeau}, {Tsang}, {Tse}, {Tso}, {Tsukada}, {Tsuna}, {Tsutsui}, {Turconi},
  {Ubhi}, {Udall}, {Ueno}, {Ugolini}, {Unnikrishnan}, {Urban}, {Usman},
  {Utina}, {Vahlbruch}, {Vajente}, {Valdes}, {Valentini}, {van Bakel}, {van
  Beuzekom}, {van den Brand}, {Van Den Broeck}, {Vander-Hyde}, {van der
  Schaaf}, {Van Heijningen}, {van Veggel}, {Vardaro}, {Varma}, {Vass},
  {Vas{\'u}th}, {Vecchio}, {Vedovato}, {Veitch}, {Veitch}, {Venkateswara},
  {Venugopalan}, {Verkindt}, {Veske}, {Vetrano}, {Vicer{\'e}}, {Viets},
  {Vinciguerra}, {Vine}, {Vinet}, {Vitale}, {Vivanco}, {Vo}, {Vocca},
  {Vorvick}, {Vyatchanin}, {Wade}, {Wade}, {Wade}, {Walet}, {Walker},
  {Wallace}, {Wallace}, {Walsh}, {Wang}, {Wang}, {Wang}, {Ward}, {Warden},
  {Warner}, {Was}, {Watchi}, {Weaver}, {Wei}, {Weinert}, {Weinstein}, {Weiss},
  {Wellmann}, {Wen}, {We{\ss}els}, {Westhouse}, {Wette}, {Whelan}, {Whiting},
  {Whittle}, {Wilken}, {Williams}, {Willis}, {Willke}, {Winkler}, {Wipf},
  {Wittel}, {Woan}, {Woehler}, {Wofford}, {Wong}, {Wright}, {Wu}, {Wysocki},
  {Xiao}, {Yamamoto}, {Yang}, {Yang}, {Yang}, {Yap}, {Yazback}, {Yeeles}, {Yu},
  {Yu}, {Yuen}, {Zadro{\.Z}ny}, {Zadro{\.Z}ny}, {Zanolin}, {Zelenova},
  {Zendri}, {Zevin}, {Zhang}, {Zhang}, {Zhang}, {Zhao}, {Zhao}, {Zhou}, {Zhou},
  {Zhu}, {Zimmerman}, {Zucker}, {Zweizig}, {LIGO Scientific Collaboration}, \&
  {Virgo Collaboration}}]{Abbott2020}
{Abbott}, R., {Abbott}, T.~D., {Abraham}, S., {et~al.} 2020, \prl, 125, 101102,
  \dodoi{10.1103/PhysRevLett.125.101102}

\bibitem[{{Amaro-Seoane} {et~al.}(2017){Amaro-Seoane}, {Audley}, {Babak},
  {Baker}, {Barausse}, {Bender}, {Berti}, {Binetruy}, {Born}, {Bortoluzzi},
  {Camp}, {Caprini}, {Cardoso}, {Colpi}, {Conklin}, {Cornish}, {Cutler},
  {Danzmann}, {Dolesi}, {Ferraioli}, {Ferroni}, {Fitzsimons}, {Gair}, {Gesa
  Bote}, {Giardini}, {Gibert}, {Grimani}, {Halloin}, {Heinzel}, {Hertog},
  {Hewitson}, {Holley-Bockelmann}, {Hollington}, {Hueller}, {Inchauspe},
  {Jetzer}, {Karnesis}, {Killow}, {Klein}, {Klipstein}, {Korsakova}, {Larson},
  {Livas}, {Lloro}, {Man}, {Mance}, {Martino}, {Mateos}, {McKenzie},
  {McWilliams}, {Miller}, {Mueller}, {Nardini}, {Nelemans}, {Nofrarias},
  {Petiteau}, {Pivato}, {Plagnol}, {Porter}, {Reiche}, {Robertson},
  {Robertson}, {Rossi}, {Russano}, {Schutz}, {Sesana}, {Shoemaker}, {Slutsky},
  {Sopuerta}, {Sumner}, {Tamanini}, {Thorpe}, {Troebs}, {Vallisneri},
  {Vecchio}, {Vetrugno}, {Vitale}, {Volonteri}, {Wanner}, {Ward}, {Wass},
  {Weber}, {Ziemer}, \& {Zweifel}}]{Amaro2017}
{Amaro-Seoane}, P., {Audley}, H., {Babak}, S., {et~al.} 2017, arXiv e-prints,
  arXiv:1702.00786, \dodoi{10.48550/arXiv.1702.00786}

\bibitem[{{Angus} {et~al.}(2022){Angus}, {Baldassare}, {Mockler}, {Foley},
  {Ramirez-Ruiz}, {Raimundo}, {French}, {Auchettl}, {Pfister}, {Gall},
  {Hjorth}, {Drout}, {Alexander}, {Dimitriadis}, {Hung}, {Jones}, {Rest},
  {Siebert}, {Taggart}, {Terreran}, {Tinyanont}, {Carroll}, {DeMarchi}, {Earl},
  {Gagliano}, {Izzo}, {Villar}, {Zenati}, {Arendse}, {Cold}, {de Boer},
  {Chambers}, {Coulter}, {Khetan}, {Lin}, {Magnier}, {Rojas-Bravo},
  {Wainscoat}, \& {Wojtak}}]{2022Angus}
{Angus}, C.~R., {Baldassare}, V.~F., {Mockler}, B., {et~al.} 2022, Nature
  Astronomy, 6, 1452, \dodoi{10.1038/s41550-022-01811-y}

\bibitem[{{Antonini} {et~al.}(2016){Antonini}, {Chatterjee}, {Rodriguez},
  {Morscher}, {Pattabiraman}, {Kalogera}, \& {Rasio}}]{Antonini2016}
{Antonini}, F., {Chatterjee}, S., {Rodriguez}, C.~L., {et~al.} 2016, \apj, 816,
  65, \dodoi{10.3847/0004-637X/816/2/65}

\bibitem[{{Antonini} \& {Gieles}(2020{\natexlab{a}})}]{Antonini2020a}
{Antonini}, F., \& {Gieles}, M. 2020{\natexlab{a}}, \prd, 102, 123016,
  \dodoi{10.1103/PhysRevD.102.123016}

\bibitem[{{Antonini} \& {Gieles}(2020{\natexlab{b}})}]{Antonini2020b}
---. 2020{\natexlab{b}}, \mnras, 492, 2936, \dodoi{10.1093/mnras/stz3584}

\bibitem[{{Antonini} {et~al.}(2019){Antonini}, {Gieles}, \&
  {Gualandris}}]{Antonini2019}
{Antonini}, F., {Gieles}, M., \& {Gualandris}, A. 2019, \mnras, 486, 5008,
  \dodoi{10.1093/mnras/stz1149}

\bibitem[{{Astropy Collaboration} {et~al.}(2018){Astropy Collaboration},
  {Price-Whelan}, {Sip{\H{o}}cz}, {G{\"u}nther}, {Lim}, {Crawford}, {Conseil},
  {Shupe}, {Craig}, {Dencheva}, {Ginsburg}, {VanderPlas}, {Bradley},
  {P{\'e}rez-Su{\'a}rez}, {de Val-Borro}, {Aldcroft}, {Cruz}, {Robitaille},
  {Tollerud}, {Ardelean}, {Babej}, {Bach}, {Bachetti}, {Bakanov}, {Bamford},
  {Barentsen}, {Barmby}, {Baumbach}, {Berry}, {Biscani}, {Boquien}, {Bostroem},
  {Bouma}, {Brammer}, {Bray}, {Breytenbach}, {Buddelmeijer}, {Burke},
  {Calderone}, {Cano Rodr{\'\i}guez}, {Cara}, {Cardoso}, {Cheedella}, {Copin},
  {Corrales}, {Crichton}, {D'Avella}, {Deil}, {Depagne}, {Dietrich}, {Donath},
  {Droettboom}, {Earl}, {Erben}, {Fabbro}, {Ferreira}, {Finethy}, {Fox},
  {Garrison}, {Gibbons}, {Goldstein}, {Gommers}, {Greco}, {Greenfield},
  {Groener}, {Grollier}, {Hagen}, {Hirst}, {Homeier}, {Horton}, {Hosseinzadeh},
  {Hu}, {Hunkeler}, {Ivezi{\'c}}, {Jain}, {Jenness}, {Kanarek}, {Kendrew},
  {Kern}, {Kerzendorf}, {Khvalko}, {King}, {Kirkby}, {Kulkarni}, {Kumar},
  {Lee}, {Lenz}, {Littlefair}, {Ma}, {Macleod}, {Mastropietro}, {McCully},
  {Montagnac}, {Morris}, {Mueller}, {Mumford}, {Muna}, {Murphy}, {Nelson},
  {Nguyen}, {Ninan}, {N{\"o}the}, {Ogaz}, {Oh}, {Parejko}, {Parley}, {Pascual},
  {Patil}, {Patil}, {Plunkett}, {Prochaska}, {Rastogi}, {Reddy Janga},
  {Sabater}, {Sakurikar}, {Seifert}, {Sherbert}, {Sherwood-Taylor}, {Shih},
  {Sick}, {Silbiger}, {Singanamalla}, {Singer}, {Sladen}, {Sooley},
  {Sornarajah}, {Streicher}, {Teuben}, {Thomas}, {Tremblay}, {Turner},
  {Terr{\'o}n}, {van Kerkwijk}, {de la Vega}, {Watkins}, {Weaver}, {Whitmore},
  {Woillez}, {Zabalza}, \& {Astropy Contributors}}]{Astropy2018}
{Astropy Collaboration}, {Price-Whelan}, A.~M., {Sip{\H{o}}cz}, B.~M., {et~al.}
  2018, \aj, 156, 123, \dodoi{10.3847/1538-3881/aabc4f}

\bibitem[{{Atallah} {et~al.}(2023){Atallah}, {Trani}, {Kremer}, {Weatherford},
  {Fragione}, {Spera}, \& {Rasio}}]{Atallah2023}
{Atallah}, D., {Trani}, A.~A., {Kremer}, K., {et~al.} 2023, \mnras, 523, 4227,
  \dodoi{10.1093/mnras/stad1634}

\bibitem[{{Banerjee} {et~al.}(2020){Banerjee}, {Belczynski}, {Fryer},
  {Berczik}, {Hurley}, {Spurzem}, \& {Wang}}]{2020Banerjee}
{Banerjee}, S., {Belczynski}, K., {Fryer}, C.~L., {et~al.} 2020, \aap, 639,
  A41, \dodoi{10.1051/0004-6361/201935332}

\bibitem[{{Begelman} {et~al.}(2006){Begelman}, {Volonteri}, \&
  {Rees}}]{Begelman2006}
{Begelman}, M.~C., {Volonteri}, M., \& {Rees}, M.~J. 2006, \mnras, 370, 289,
  \dodoi{10.1111/j.1365-2966.2006.10467.x}

\bibitem[{{Bromm} \& {Loeb}(2003)}]{Bromm2003}
{Bromm}, V., \& {Loeb}, A. 2003, \apj, 596, 34, \dodoi{10.1086/377529}

\bibitem[{{Celotti} {et~al.}(1999){Celotti}, {Miller}, \&
  {Sciama}}]{Celotti1999}
{Celotti}, A., {Miller}, J.~C., \& {Sciama}, D.~W. 1999, Classical and Quantum
  Gravity, 16, A3, \dodoi{10.1088/0264-9381/16/12A/301}

\bibitem[{{Chattopadhyay} {et~al.}(2023){Chattopadhyay}, {Stegmann},
  {Antonini}, {Barber}, \& {Romero-Shaw}}]{Chattopadhyay2023}
{Chattopadhyay}, D., {Stegmann}, J., {Antonini}, F., {Barber}, J., \&
  {Romero-Shaw}, I.~M. 2023, \mnras, 526, 4908, \dodoi{10.1093/mnras/stad3048}

\bibitem[{{Di Carlo} {et~al.}(2021){Di Carlo}, {Mapelli}, {Pasquato},
  {Rastello}, {Ballone}, {Dall'Amico}, {Giacobbo}, {Iorio}, {Spera},
  {Torniamenti}, \& {Haardt}}]{DiCarlo2021}
{Di Carlo}, U.~N., {Mapelli}, M., {Pasquato}, M., {et~al.} 2021, \mnras, 507,
  5132, \dodoi{10.1093/mnras/stab2390}

\bibitem[{{Fishbach} \& {Fragione}(2023)}]{Fishbach2023}
{Fishbach}, M., \& {Fragione}, G. 2023, \mnras, 522, 5546,
  \dodoi{10.1093/mnras/stad1364}

\bibitem[{{Fragione}(2022)}]{Fragione2022}
{Fragione}, G. 2022, \apj, 939, 97, \dodoi{10.3847/1538-4357/ac98b6}

\bibitem[{{Fragione}(2023)}]{2023zndoFragione}
---. 2023, {giacomofragione/imbhistory: imbhistory v1.01}, v1.01, Zenodo,
  Zenodo, \dodoi{10.5281/zenodo.7530024}

\bibitem[{{Fragione} \& {Loeb}(2023)}]{FragioneLoeb2023}
{Fragione}, G., \& {Loeb}, A. 2023, \apj, 944, 81,
  \dodoi{10.3847/1538-4357/acb34e}

\bibitem[{{Fragione} {et~al.}(2021){Fragione}, {Perna}, \&
  {Loeb}}]{Fragione2021}
{Fragione}, G., {Perna}, R., \& {Loeb}, A. 2021, \mnras, 500, 4307,
  \dodoi{10.1093/mnras/staa3493}

\bibitem[{{Fragione} \& {Rasio}(2023)}]{FragioneRasio2023}
{Fragione}, G., \& {Rasio}, F.~A. 2023, \apj, 951, 129,
  \dodoi{10.3847/1538-4357/acd9c9}

\bibitem[{{Fregeau} {et~al.}(2002){Fregeau}, {Joshi}, {Portegies Zwart}, \&
  {Rasio}}]{Fregeau2002}
{Fregeau}, J.~M., {Joshi}, K.~J., {Portegies Zwart}, S.~F., \& {Rasio}, F.~A.
  2002, \apj, 570, 171, \dodoi{10.1086/339576}

\bibitem[{{Freitag}(2001)}]{Freitag2001}
{Freitag}, M. 2001, Classical and Quantum Gravity, 18, 4033,
  \dodoi{10.1088/0264-9381/18/19/309}

\bibitem[{{Fryer} \& {Kalogera}(2001)}]{FryerKalogera2001}
{Fryer}, C.~L., \& {Kalogera}, V. 2001, \apj, 554, 548, \dodoi{10.1086/321359}

\bibitem[{{Fuller} \& {Ma}(2019)}]{FullerMa2019}
{Fuller}, J., \& {Ma}, L. 2019, \apjl, 881, L1,
  \dodoi{10.3847/2041-8213/ab339b}

\bibitem[{{Giersz} {et~al.}(2015{\natexlab{a}}){Giersz}, {Leigh}, {Hypki},
  {L{\"u}tzgendorf}, \& {Askar}}]{Giersz2015}
{Giersz}, M., {Leigh}, N., {Hypki}, A., {L{\"u}tzgendorf}, N., \& {Askar}, A.
  2015{\natexlab{a}}, \mnras, 454, 3150, \dodoi{10.1093/mnras/stv2162}

\bibitem[{{Giersz} {et~al.}(2015{\natexlab{b}}){Giersz}, {Leigh}, {Hypki},
  {L{\"u}tzgendorf}, \& et~al.}]{GierszLeigh2015}
{Giersz}, M., {Leigh}, N., {Hypki}, A., {L{\"u}tzgendorf}, N., \& et~al.
  2015{\natexlab{b}}, \mnras, 454, 3150, \dodoi{10.1093/mnras/stv2162}

\bibitem[{{Gonz{\'a}lez} {et~al.}(2021){Gonz{\'a}lez}, {Kremer}, {Chatterjee},
  {Fragione}, {Rodriguez}, {Weatherford}, {Ye}, \& {Rasio}}]{Gonzalez2021}
{Gonz{\'a}lez}, E., {Kremer}, K., {Chatterjee}, S., {et~al.} 2021, \apjl, 908,
  L29, \dodoi{10.3847/2041-8213/abdf5b}

\bibitem[{{Graham} \& {Scott}(2013)}]{GrahamScott2013}
{Graham}, A.~W., \& {Scott}, N. 2013, \apj, 764, 151,
  \dodoi{10.1088/0004-637X/764/2/151}

\bibitem[{{Greene} {et~al.}(2020){Greene}, {Strader}, \& {Ho}}]{Greene2020}
{Greene}, J.~E., {Strader}, J., \& {Ho}, L.~C. 2020, \araa, 58, 257,
  \dodoi{10.1146/annurev-astro-032620-021835}

\bibitem[{{G{\"u}rkan} {et~al.}(2006){G{\"u}rkan}, {Fregeau}, \&
  {Rasio}}]{Gurkan2006}
{G{\"u}rkan}, M.~A., {Fregeau}, J.~M., \& {Rasio}, F.~A. 2006, \apjl, 640, L39,
  \dodoi{10.1086/503295}

\bibitem[{{G{\"u}rkan} {et~al.}(2004){G{\"u}rkan}, {Freitag}, \&
  {Rasio}}]{Gurkan2004}
{G{\"u}rkan}, M.~A., {Freitag}, M., \& {Rasio}, F.~A. 2004, \apj, 604, 632,
  \dodoi{10.1086/381968}

\bibitem[{{Heger} {et~al.}(2003){Heger}, {Fryer}, {Woosley}, {Langer}, \&
  {Hartmann}}]{Heger2003}
{Heger}, A., {Fryer}, C.~L., {Woosley}, S.~E., {Langer}, N., \& {Hartmann},
  D.~H. 2003, \apj, 591, 288, \dodoi{10.1086/375341}

\bibitem[{{Hobbs} {et~al.}(2005){Hobbs}, {Lorimer}, {Lyne}, \&
  {Kramer}}]{Hobbs2005}
{Hobbs}, G., {Lorimer}, D.~R., {Lyne}, A.~G., \& {Kramer}, M. 2005, \mnras,
  360, 974, \dodoi{10.1111/j.1365-2966.2005.09087.x}

\bibitem[{{Hurley} {et~al.}(2000){Hurley}, {Pols}, \& {Tout}}]{Hurley2000}
{Hurley}, J.~R., {Pols}, O.~R., \& {Tout}, C.~A. 2000, \mnras, 315, 543,
  \dodoi{10.1046/j.1365-8711.2000.03426.x}

\bibitem[{{Husa} {et~al.}(2016){Husa}, {Khan}, {Hannam}, {P{\"u}rrer}, {Ohme},
  {Forteza}, \& {Boh{\'e}}}]{Husa2016}
{Husa}, S., {Khan}, S., {Hannam}, M., {et~al.} 2016, \prd, 93, 044006,
  \dodoi{10.1103/PhysRevD.93.044006}

\bibitem[{{Ivanova} {et~al.}(2005){Ivanova}, {Belczynski}, {Fregeau}, \&
  {Rasio}}]{Ivanova2005}
{Ivanova}, N., {Belczynski}, K., {Fregeau}, J.~M., \& {Rasio}, F.~A. 2005,
  \mnras, 358, 572, \dodoi{10.1111/j.1365-2966.2005.08804.x}

\bibitem[{{Jani} {et~al.}(2020){Jani}, {Shoemaker}, \& {Cutler}}]{Jani2020}
{Jani}, K., {Shoemaker}, D., \& {Cutler}, C. 2020, Nature Astronomy, 4, 260,
  \dodoi{10.1038/s41550-019-0932-7}

\bibitem[{{Jardel} \& {Gebhardt}(2012)}]{Jardel2012}
{Jardel}, J.~R., \& {Gebhardt}, K. 2012, \apj, 746, 89,
  \dodoi{10.1088/0004-637X/746/1/89}

\bibitem[{{Jiang} {et~al.}(2019){Jiang}, {Stone}, \& {Davis}}]{Jiang2019}
{Jiang}, Y.-F., {Stone}, J.~M., \& {Davis}, S.~W. 2019, \apj, 880, 67,
  \dodoi{10.3847/1538-4357/ab29ff}

\bibitem[{{King}(1962)}]{King1962}
{King}, I. 1962, \aj, 67, 471, \dodoi{10.1086/108756}

\bibitem[{{K{\i}ro{\u{g}}lu} {et~al.}(2023){K{\i}ro{\u{g}}lu}, {Lombardi},
  {Kremer}, {Fragione}, {Fogarty}, \& {Rasio}}]{Fulya2023}
{K{\i}ro{\u{g}}lu}, F., {Lombardi}, J.~C., {Kremer}, K., {et~al.} 2023, \apj,
  948, 89, \dodoi{10.3847/1538-4357/acc24c}

\bibitem[{{Kormendy} \& {Ho}(2013)}]{Kormendy2013}
{Kormendy}, J., \& {Ho}, L.~C. 2013, \araa, 51, 511,
  \dodoi{10.1146/annurev-astro-082708-101811}

\bibitem[{{Kroupa}(2001)}]{Kroupa2001}
{Kroupa}, P. 2001, \mnras, 322, 231, \dodoi{10.1046/j.1365-8711.2001.04022.x}

\bibitem[{{Liu} {et~al.}(2009){Liu}, {Li}, \& {Chen}}]{Liu2009}
{Liu}, F.~K., {Li}, S., \& {Chen}, X. 2009, \apjl, 706, L133,
  \dodoi{10.1088/0004-637X/706/1/L133}

\bibitem[{{Lousto} {et~al.}(2010){Lousto}, {Campanelli}, {Zlochower}, \&
  {Nakano}}]{LoustoCampanelli2010}
{Lousto}, C.~O., {Campanelli}, M., {Zlochower}, Y., \& {Nakano}, H. 2010,
  Classical and Quantum Gravity, 27, 114006,
  \dodoi{10.1088/0264-9381/27/11/114006}

\bibitem[{{Madau} \& {Fragos}(2017)}]{MadauFragos2017}
{Madau}, P., \& {Fragos}, T. 2017, \apj, 840, 39,
  \dodoi{10.3847/1538-4357/aa6af9}

\bibitem[{{Madau} \& {Rees}(2001)}]{Madau2001}
{Madau}, P., \& {Rees}, M.~J. 2001, \apjl, 551, L27, \dodoi{10.1086/319848}

\bibitem[{{Mapelli}(2016)}]{Mapeli2016}
{Mapelli}, M. 2016, \mnras, 459, 3432, \dodoi{10.1093/mnras/stw869}

\bibitem[{{Mapelli} {et~al.}(2021){Mapelli}, {Dall'Amico}, {Bouffanais},
  {Giacobbo}, {Arca Sedda}, {Artale}, {Ballone}, {Di Carlo}, {Iorio},
  {Santoliquido}, \& {Torniamenti}}]{Mapelli2021}
{Mapelli}, M., {Dall'Amico}, M., {Bouffanais}, Y., {et~al.} 2021, \mnras, 505,
  339, \dodoi{10.1093/mnras/stab1334}

\bibitem[{{Marconi} \& {Hunt}(2003)}]{Marconi2003}
{Marconi}, A., \& {Hunt}, L.~K. 2003, \apjl, 589, L21, \dodoi{10.1086/375804}

\bibitem[{{Mezcua}(2017)}]{Mezcua2017}
{Mezcua}, M. 2017, International Journal of Modern Physics D, 26, 1730021,
  \dodoi{10.1142/S021827181730021X}

\bibitem[{{Miller} \& {Hamilton}(2002)}]{Miller2002}
{Miller}, M.~C., \& {Hamilton}, D.~P. 2002, \mnras, 330, 232,
  \dodoi{10.1046/j.1365-8711.2002.05112.x}

\bibitem[{{Neumayer} \& {Walcher}(2012)}]{Neumayer2012}
{Neumayer}, N., \& {Walcher}, C.~J. 2012, Advances in Astronomy, 2012, 709038,
  \dodoi{10.1155/2012/709038}

\bibitem[{{Nguyen} {et~al.}(2018){Nguyen}, {Seth}, {Neumayer}, {Kamann},
  {Voggel}, {Cappellari}, {Picotti}, {Nguyen}, {B{\"o}ker}, {Debattista},
  {Caldwell}, {McDermid}, {Bastian}, {Ahn}, \& {Pechetti}}]{Nguyen2018}
{Nguyen}, D.~D., {Seth}, A.~C., {Neumayer}, N., {et~al.} 2018, \apj, 858, 118,
  \dodoi{10.3847/1538-4357/aabe28}

\bibitem[{{Nitz} {et~al.}(2019){Nitz}, {Harry}, {Brown}, {Biwer}, {Willis},
  {Dal Canton}, {Capano}, {Pekowsky}, {Dent}, {Williamson}, {De}, {Davies},
  {Cabero}, {Macleod}, {Machenschalk}, {Reyes}, {Kumar}, {Massinger},
  {Pannarale}, {Dfinstad}, {T{\'a}pai}, {Fairhurst}, {Khan}, {Singer},
  {Nielsen}, {Shasvath}, {Kumar}, {Idorrington92}, {Gabbard}, \& {Uday Varsha
  Gadre}}]{Nitz2019}
{Nitz}, A., {Harry}, I., {Brown}, D., {et~al.} 2019, {gwastro/pycbc: PyCBC
  Release v1.15.2}, v1.15.2, Zenodo,  Zenodo, \dodoi{10.5281/zenodo.3564824}

\bibitem[{{Perley} {et~al.}(2019){Perley}, {Mazzali}, {Yan}, {Cenko}, \&
  et~al.}]{PerleyMazzali2019}
{Perley}, D.~A., {Mazzali}, P.~A., {Yan}, L., {Cenko}, S.~B., \& et~al. 2019,
  \mnras, 484, 1031, \dodoi{10.1093/mnras/sty3420}

\bibitem[{{Planck Collaboration} {et~al.}(2016){Planck Collaboration}, {Ade},
  {Aghanim}, {Arnaud}, {Ashdown}, {Aumont}, {Baccigalupi}, {Banday},
  {Barreiro}, {Bartlett}, {Bartolo}, {Battaner}, {Battye}, {Benabed},
  {Beno{\^\i}t}, {Benoit-L{\'e}vy}, {Bernard}, {Bersanelli}, {Bielewicz},
  {Bock}, {Bonaldi}, {Bonavera}, {Bond}, {Borrill}, {Bouchet}, {Boulanger},
  {Bucher}, {Burigana}, {Butler}, {Calabrese}, {Cardoso}, {Catalano},
  {Challinor}, {Chamballu}, {Chary}, {Chiang}, {Chluba}, {Christensen},
  {Church}, {Clements}, {Colombi}, {Colombo}, {Combet}, {Coulais}, {Crill},
  {Curto}, {Cuttaia}, {Danese}, {Davies}, {Davis}, {de Bernardis}, {de Rosa},
  {de Zotti}, {Delabrouille}, {D{\'e}sert}, {Di Valentino}, {Dickinson},
  {Diego}, {Dolag}, {Dole}, {Donzelli}, {Dor{\'e}}, {Douspis}, {Ducout},
  {Dunkley}, {Dupac}, {Efstathiou}, {Elsner}, {En{\ss}lin}, {Eriksen},
  {Farhang}, {Fergusson}, {Finelli}, {Forni}, {Frailis}, {Fraisse},
  {Franceschi}, {Frejsel}, {Galeotta}, {Galli}, {Ganga}, {Gauthier}, {Gerbino},
  {Ghosh}, {Giard}, {Giraud-H{\'e}raud}, {Giusarma}, {Gjerl{\o}w},
  {Gonz{\'a}lez-Nuevo}, {G{\'o}rski}, {Gratton}, {Gregorio}, {Gruppuso},
  {Gudmundsson}, {Hamann}, {Hansen}, {Hanson}, {Harrison}, {Helou},
  {Henrot-Versill{\'e}}, {Hern{\'a}ndez-Monteagudo}, {Herranz}, {Hildebrandt},
  {Hivon}, {Hobson}, {Holmes}, {Hornstrup}, {Hovest}, {Huang}, {Huffenberger},
  {Hurier}, {Jaffe}, {Jaffe}, {Jones}, {Juvela}, {Keih{\"a}nen}, {Keskitalo},
  {Kisner}, {Kneissl}, {Knoche}, {Knox}, {Kunz}, {Kurki-Suonio}, {Lagache},
  {L{\"a}hteenm{\"a}ki}, {Lamarre}, {Lasenby}, {Lattanzi}, {Lawrence}, {Leahy},
  {Leonardi}, {Lesgourgues}, {Levrier}, {Lewis}, {Liguori}, {Lilje},
  {Linden-V{\o}rnle}, {L{\'o}pez-Caniego}, {Lubin}, {Mac{\'\i}as-P{\'e}rez},
  {Maggio}, {Maino}, {Mandolesi}, {Mangilli}, {Marchini}, {Maris}, {Martin},
  {Martinelli}, {Mart{\'\i}nez-Gonz{\'a}lez}, {Masi}, {Matarrese}, {McGehee},
  {Meinhold}, {Melchiorri}, {Melin}, {Mendes}, {Mennella}, {Migliaccio},
  {Millea}, {Mitra}, {Miville-Desch{\^e}nes}, {Moneti}, {Montier}, {Morgante},
  {Mortlock}, {Moss}, {Munshi}, {Murphy}, {Naselsky}, {Nati}, {Natoli},
  {Netterfield}, {N{\o}rgaard-Nielsen}, {Noviello}, {Novikov}, {Novikov},
  {Oxborrow}, {Paci}, {Pagano}, {Pajot}, {Paladini}, {Paoletti}, {Partridge},
  {Pasian}, {Patanchon}, {Pearson}, {Perdereau}, {Perotto}, {Perrotta},
  {Pettorino}, {Piacentini}, {Piat}, {Pierpaoli}, {Pietrobon}, {Plaszczynski},
  {Pointecouteau}, {Polenta}, {Popa}, {Pratt}, {Pr{\'e}zeau}, {Prunet},
  {Puget}, {Rachen}, {Reach}, {Rebolo}, {Reinecke}, {Remazeilles}, {Renault},
  {Renzi}, {Ristorcelli}, {Rocha}, {Rosset}, {Rossetti}, {Roudier},
  {Rouill{\'e} d'Orfeuil}, {Rowan-Robinson}, {Rubi{\~n}o-Mart{\'\i}n},
  {Rusholme}, {Said}, {Salvatelli}, {Salvati}, {Sandri}, {Santos},
  {Savelainen}, {Savini}, {Scott}, {Seiffert}, {Serra}, {Shellard}, {Spencer},
  {Spinelli}, {Stolyarov}, {Stompor}, {Sudiwala}, {Sunyaev}, {Sutton},
  {Suur-Uski}, {Sygnet}, {Tauber}, {Terenzi}, {Toffolatti}, {Tomasi},
  {Tristram}, {Trombetti}, {Tucci}, {Tuovinen}, {T{\"u}rler}, {Umana},
  {Valenziano}, {Valiviita}, {Van Tent}, {Vielva}, {Villa}, {Wade}, {Wandelt},
  {Wehus}, {White}, {White}, {Wilkinson}, {Yvon}, {Zacchei}, \&
  {Zonca}}]{Planck2016}
{Planck Collaboration}, {Ade}, P.~A.~R., {Aghanim}, N., {et~al.} 2016, \aap,
  594, A13, \dodoi{10.1051/0004-6361/201525830}

\bibitem[{{Portegies Zwart} {et~al.}(2004){Portegies Zwart}, {Baumgardt},
  {Hut}, {Makino}, \& {McMillan}}]{Zwart2004}
{Portegies Zwart}, S.~F., {Baumgardt}, H., {Hut}, P., {Makino}, J., \&
  {McMillan}, S. L.~W. 2004, \nat, 428, 724, \dodoi{10.1038/nature02448}

\bibitem[{{Portegies Zwart} \& {McMillan}(2000)}]{Portegies2000}
{Portegies Zwart}, S.~F., \& {McMillan}, S. L.~W. 2000, \apjl, 528, L17,
  \dodoi{10.1086/312422}

\bibitem[{{Portegies Zwart} \& {McMillan}(2002)}]{Portegies2002}
---. 2002, \apj, 576, 899, \dodoi{10.1086/341798}

\bibitem[{{Portegies Zwart} {et~al.}(2010){Portegies Zwart}, {McMillan}, \&
  {Gieles}}]{Portegies2010}
{Portegies Zwart}, S.~F., {McMillan}, S. L.~W., \& {Gieles}, M. 2010, \araa,
  48, 431, \dodoi{10.1146/annurev-astro-081309-130834}

\bibitem[{{Punturo} {et~al.}(2010){Punturo}, {Abernathy}, {Acernese}, {Allen},
  {Andersson}, {Arun}, {Barone}, {Barr}, {Barsuglia}, {Beker}, {Beveridge},
  {Birindelli}, {Bose}, {Bosi}, {Braccini}, {Bradaschia}, {Bulik}, {Calloni},
  {Cella}, {Chassande Mottin}, {Chelkowski}, {Chincarini}, {Clark}, {Coccia},
  {Colacino}, {Colas}, {Cumming}, {Cunningham}, {Cuoco}, {Danilishin},
  {Danzmann}, {De Luca}, {De Salvo}, {Dent}, {De Rosa}, {Di Fiore}, {Di
  Virgilio}, {Doets}, {Fafone}, {Falferi}, {Flaminio}, {Franc}, {Frasconi},
  {Freise}, {Fulda}, {Gair}, {Gemme}, {Gennai}, {Giazotto}, {Glampedakis},
  {Granata}, {Grote}, {Guidi}, {Hammond}, {Hannam}, {Harms}, {Heinert},
  {Hendry}, {Heng}, {Hennes}, {Hild}, {Hough}, {Husa}, {Huttner}, {Jones},
  {Khalili}, {Kokeyama}, {Kokkotas}, {Krishnan}, {Lorenzini}, {L{\"u}ck},
  {Majorana}, {Mandel}, {Mandic}, {Martin}, {Michel}, {Minenkov}, {Morgado},
  {Mosca}, {Mours}, {M{\"u}ller{\textendash}Ebhardt}, {Murray}, {Nawrodt},
  {Nelson}, {Oshaughnessy}, {Ott}, {Palomba}, {Paoli}, {Parguez},
  {Pasqualetti}, {Passaquieti}, {Passuello}, {Pinard}, {Poggiani}, {Popolizio},
  {Prato}, {Puppo}, {Rabeling}, {Rapagnani}, {Read}, {Regimbau}, {Rehbein},
  {Reid}, {Rezzolla}, {Ricci}, {Richard}, {Rocchi}, {Rowan}, {R{\"u}diger},
  {Sassolas}, {Sathyaprakash}, {Schnabel}, {Schwarz}, {Seidel}, {Sintes},
  {Somiya}, {Speirits}, {Strain}, {Strigin}, {Sutton}, {Tarabrin},
  {Th{\"u}ring}, {van den Brand}, {van Leewen}, {van Veggel}, {van den Broeck},
  {Vecchio}, {Veitch}, {Vetrano}, {Vicere}, {Vyatchanin}, {Willke}, {Woan},
  {Wolfango}, \& {Yamamoto}}]{Punturo2010}
{Punturo}, M., {Abernathy}, M., {Acernese}, F., {et~al.} 2010, Classical and
  Quantum Gravity, 27, 194002, \dodoi{10.1088/0264-9381/27/19/194002}

\bibitem[{{Reitze} {et~al.}(2019){Reitze}, {Adhikari}, {Ballmer}, {Barish},
  {Barsotti}, {Billingsley}, {Brown}, {Chen}, {Coyne}, {Eisenstein}, {Evans},
  {Fritschel}, {Hall}, {Lazzarini}, {Lovelace}, {Read}, {Sathyaprakash},
  {Shoemaker}, {Smith}, {Torrie}, {Vitale}, {Weiss}, {Wipf}, \&
  {Zucker}}]{Reitze2019}
{Reitze}, D., {Adhikari}, R.~X., {Ballmer}, S., {et~al.} 2019, in Bulletin of
  the American Astronomical Society, Vol.~51, 35,
  \dodoi{10.48550/arXiv.1907.04833}

\bibitem[{{Remillard} \& {McClintock}(2006)}]{Remillard2006}
{Remillard}, R.~A., \& {McClintock}, J.~E. 2006, \araa, 44, 49,
  \dodoi{10.1146/annurev.astro.44.051905.092532}

\bibitem[{{Robson} {et~al.}(2019){Robson}, {Cornish}, \& {Liu}}]{Robson2019}
{Robson}, T., {Cornish}, N.~J., \& {Liu}, C. 2019, Classical and Quantum
  Gravity, 36, 105011, \dodoi{10.1088/1361-6382/ab1101}

\bibitem[{{Smith} {et~al.}(2023){Smith}, {Magno}, \&
  {Tripathi}}]{SmithMagno2023}
{Smith}, K.~L., {Magno}, M., \& {Tripathi}, A. 2023, \apj, 956, 3,
  \dodoi{10.3847/1538-4357/acf4f8}

\bibitem[{{The LIGO Scientific collaboration}(2019)}]{2019LIGOfuture}
{The LIGO Scientific collaboration}. 2019, arXiv e-prints, arXiv:1904.03187,
  \dodoi{10.48550/arXiv.1904.03187}

\bibitem[{{The LIGO Scientific Collaboration} {et~al.}(2021){The LIGO
  Scientific Collaboration}, {the Virgo Collaboration}, {the KAGRA
  Collaboration}, {Abbott}, \&
  et~al.}]{TheLIGOScientificCollaborationtheVirgoCollaboration2021}
{The LIGO Scientific Collaboration}, {the Virgo Collaboration}, {the KAGRA
  Collaboration}, {Abbott}, R., \& et~al. 2021, arXiv e-prints,
  arXiv:2111.03606, \dodoi{10.48550/arXiv.2111.03606}

\bibitem[{{Tremaine} {et~al.}(2002){Tremaine}, {Gebhardt}, {Bender}, {Bower},
  {Dressler}, {Faber}, {Filippenko}, {Green}, {Grillmair}, {Ho}, {Kormendy},
  {Lauer}, {Magorrian}, {Pinkney}, \& {Richstone}}]{Tremaine2002}
{Tremaine}, S., {Gebhardt}, K., {Bender}, R., {et~al.} 2002, \apj, 574, 740,
  \dodoi{10.1086/341002}

\bibitem[{Virtanen {et~al.}(2020)Virtanen, Gommers, Oliphant, Haberland, Reddy,
  Cournapeau, Burovski, Peterson, Weckesser, Bright, {van der Walt}, Brett,
  Wilson, Millman, Mayorov, Nelson, Jones, Kern, Larson, Carey, Polat, Feng,
  Moore, {VanderPlas}, Laxalde, Perktold, Cimrman, Henriksen, Quintero, Harris,
  Archibald, Ribeiro, Pedregosa, {van Mulbregt}, \& {SciPy 1.0
  Contributors}}]{2020SciPy-NMeth}
Virtanen, P., Gommers, R., Oliphant, T.~E., {et~al.} 2020, Nature Methods, 17,
  261, \dodoi{10.1038/s41592-019-0686-2}

\bibitem[{{Whalen} \& {Fryer}(2012)}]{Whalen2012}
{Whalen}, D.~J., \& {Fryer}, C.~L. 2012, \apjl, 756, L19,
  \dodoi{10.1088/2041-8205/756/1/L19}

\bibitem[{{Woosley}(2017)}]{Woosley2017}
{Woosley}, S.~E. 2017, \apj, 836, 244, \dodoi{10.3847/1538-4357/836/2/244}

\end{thebibliography}

\section*{Appendix}
\label{sec:appendix}
\begin{figure*}[htbp!]
    \centering
    \includegraphics[width=0.85\textwidth]{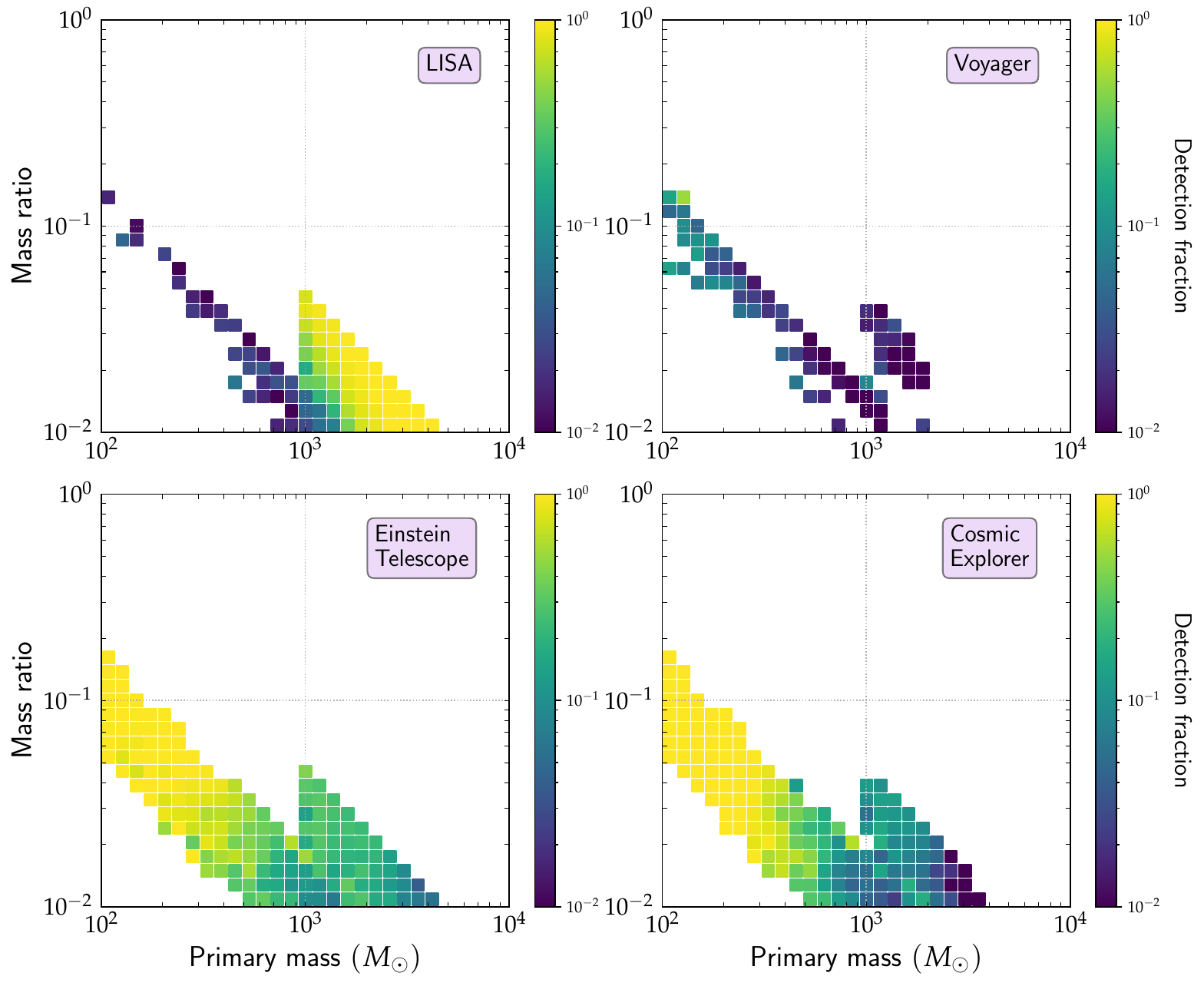}
    \caption{Same as Figure \ref{fig:gw_001} but for \frun = 0.01.}
    \label{fig:gw_01}
\end{figure*}

Here we show plots of the detection fractions by future- and ground-based GW observatories, similar to Figure~\ref{fig:gw_01}, for values of \frun = 0, 0.001 and 0.005.

\begin{figure*}[htbp!]
    \centering
    \includegraphics[width=0.85\textwidth]{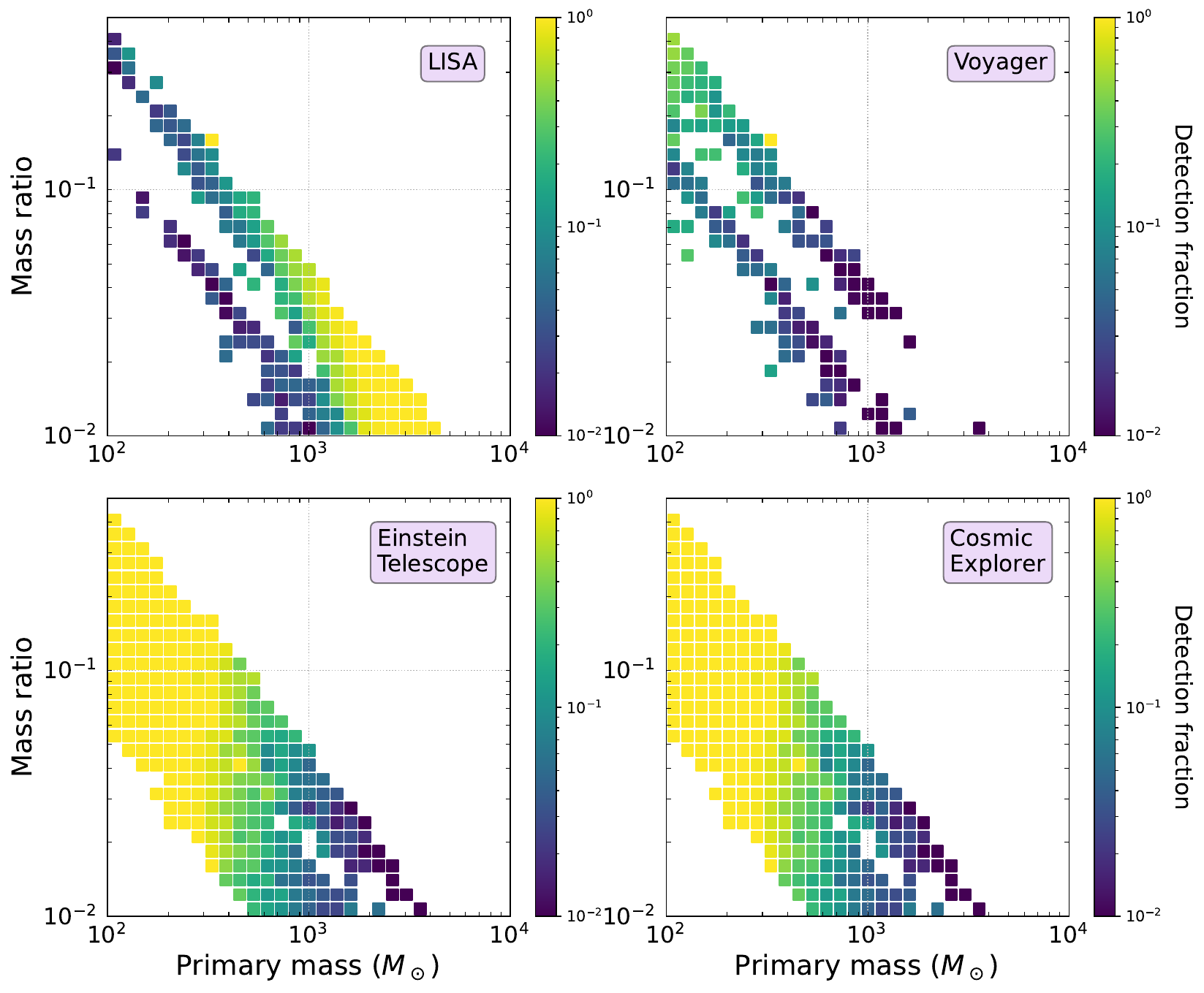}
    \caption{Same as Figure \ref{fig:gw_01} but for \frun = 0.} 
    \label{fig:gw_0}
\end{figure*}
\begin{figure*}[htbp!]
    \centering
    \includegraphics[width=0.85\textwidth]{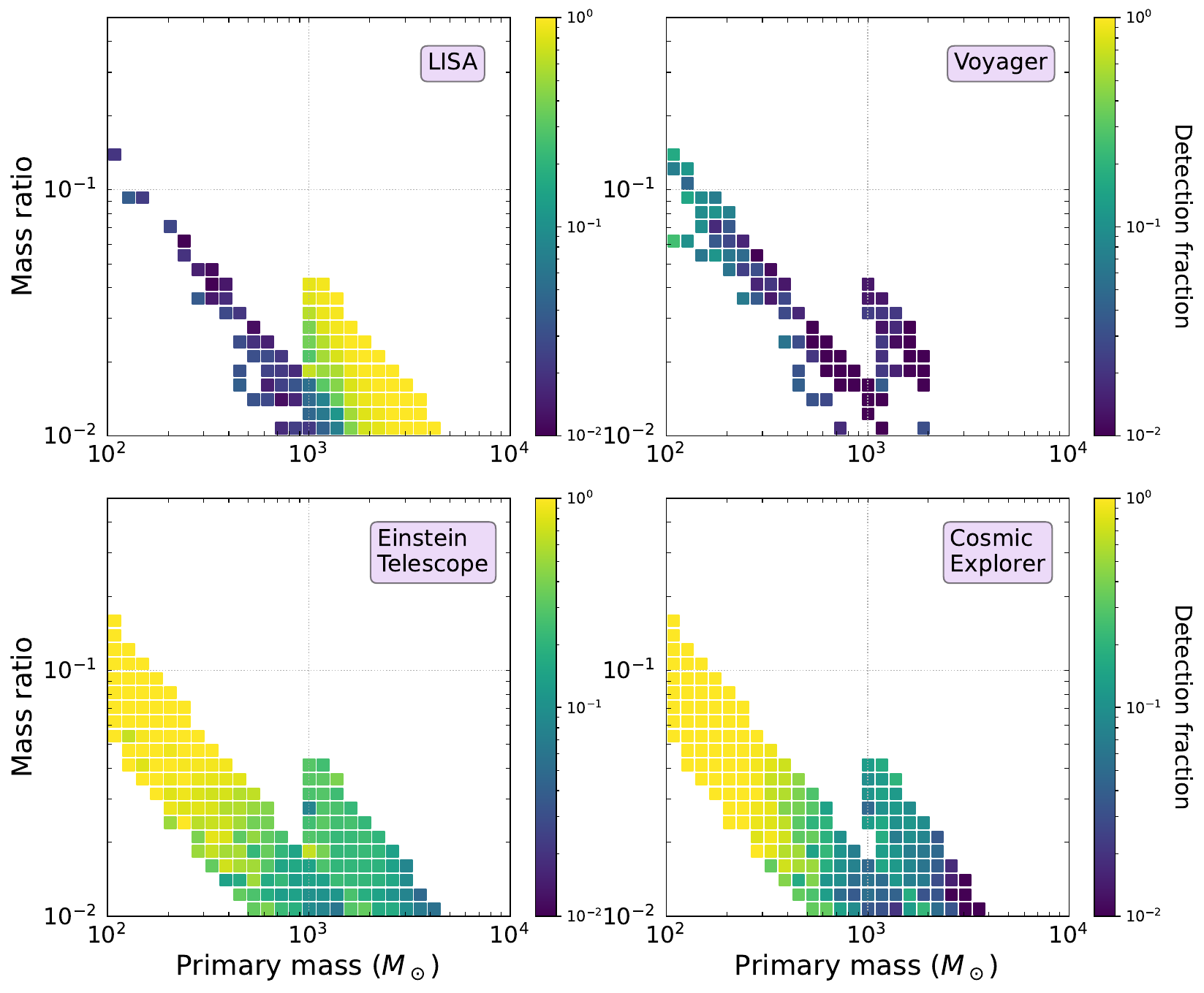}
    \caption{Same as Figure \ref{fig:gw_001} but for \frun = 0.005.}
    \label{fig:gw_005}
\end{figure*}

\end{document}